\documentclass[prc,nofootinbib]{revtex4}

\usepackage{graphicx}
\usepackage[usenames,dvipsnames]{color}
\usepackage{amsmath,amssymb,bbold}
\usepackage{hyperref}
\usepackage{comment}
\usepackage{slashed}
\usepackage{mathrsfs}
\usepackage{bbm}
\begin{document}

\title{Chiral susceptibility in Nambu Jona Lasinio model: a Wigner function approach}

\author{Arpan Das$^{1}$}
\email{arpan@prl.res.in}
\author{Deepak Kumar$^{1,2}$}
\email{deepakk@prl.res.in}
\author{Hiranmaya Mishra$^{1}$}
\email{hm@prl.res.in}

\affiliation{$^{1}$Theory Division, Physical Research Laboratory, 
Navrangpura, Ahmedabad 380 009, India}
\affiliation{$^2$ Indian Institute of Technology Gandhinagar,
	Gandhinagar 382 355, Gujarat, India}

\begin{abstract}
We estimate here chiral susceptibility at finite temperature within the framework of the Nambu-Jona-Lasinio model (NJL)
using the Wigner function approach.
We also estimate it in the presence of chiral chemical potential ($\mu_5$) as well as a non vanishing magnetic field ($B$).
We use medium separation regularization scheme (MSS) to calculate the chiral condensate and corresponding susceptibility.
It is observed that for a fixed value of chiral chemical potential ($\mu_5$), transition temperature increases with the magnetic field. While for the fixed value of the magnetic field, transition temperature decreases with chiral chemical potential. 
For a strong magnetic field, we observe non degeneracy in susceptibility for up and down type quarks.
\end{abstract}

\maketitle

\section{INTRODUCTION}

\label{intro}
In recent years extensive efforts have been made to create and understand strongly interacting matter in
relativistic heavy ion collision experiments e.g. at relativistic heavy ion collider (RHIC) and at large hadron collider (LHC).
There are mounting evidences which indicate formation of deconfined quark gluon plasma (QGP) phase of quantum chromodynamics (QCD) in the 
initial stages of these experiments as well as the formation of confined hadron phase in the subsequent evolution of QGP. Ground state of 
QCD exhibits two main non perturbative features, color confinement and  spontaneous breaking of chiral symmetry.
The dynamical breaking of chiral symmetry is the manifestation of the quark-antiquark condensation in the QCD vacuum.
Dynamical chiral symmetry breaking characterizes the non perturbative nature of QCD vacuum at vanishing temperature and/or density.
With increase in temperature and/or baryon density, the QCD vacuum undergoes a transition from a chiral symmetry broken phase to
a chiral symmetric phase. This transition is characterized by the quark-antiquark scalar condensate, the order parameter of the 
chiral phase transition. Although for first order phase transition order parameter changes discontinuously across the transition
point, for second order phase transition or for a cross over transition the variation of order parameter across the transition point is 
rather smooth. In these cases the fluctuation of this order parameter and the associated susceptibilities are more relevant for the 
characterization of the thermodynamic properties of the system.

The characteristics of fluctuations and correlations are intimately connected to the phase transition dynamics, e.g. fluctuations
of all length scales are relevant at QCD critical point where the first order quark-hadron phase transition line ends.
The study of fluctuations and correlations are essential phenomenological tool for the experimental exploration of the QCD phase diagram. 
In the context of heavy-ion collisions by studying the net electric charge fluctuation, it has been demonstrated that 
net electric charges are suppressed in the QGP phase as compared to the hadronic phase \cite{chargefluc1,chargefluc2}.  
It has also been pointed out that the correlation between baryon number and strangeness is stronger in the QGP phase as compared to the
hadronic phase \cite{strangenessenhancement1,strangenessenhancement2}. 
The quantity of interest here is the chiral susceptibility which measures the response of the chiral condensate 
to the variation of the current quark mass. Chiral susceptibility has been calculated using first principle lattice QCD (LQCD)
simulations \cite{lattice1,lattice2,lattice3,lattice4,lattice5,lattice6}. All these lattice results show a pronounced peak in the variation 
of chiral susceptibility with temperature at the transition temperature, which essentially characterizes the chiral transition. 
Apart from these LQCD studies which incorporates the non perturbative effects of QCD vacuum, complementary approaches e.g. Nambu-Jona-Lasinio
(NJL) model \cite{NJL1,NJL2}, chiral perturbation theory \cite{chipt1}, Dyson-Schwinger equation (DSE) \cite{dse1}, 
hard thermal loop approximation (HTL)\cite{htl1} etc. have been considered to study the chiral susceptibility. 

An entirely new line of investigations have been initiated to understand the QCD phase diagram
due to possibility of generation  of extremely large 
magnetic field in non central relativistic heavy ion collision experiments. In the early stages the magnetic field in QGP can be very large, at least of the 
order of few $m_{\pi}^2$ \cite{mag1,mag2,mag3,mag4,mag5,cond1,cond2,cond3,cond4}. While such fields rapidly decay in the vacuum, in a conducting medium 
they can be sustained for a longer time due to induced current \cite{cond1,cond2,cond3,cond4}.
 Strong magnetic field can affect dynamical chiral symmetry breaking. It has been shown  that external magnetic field 
act as catalysis for chiral condensation, the value of chiral condensation or the constituent mass of quarks is larger than vanishing
magnetic field case. It is important to mention that effect of magnetic field on the order parameter is not unique to QCD medium. In fact in
condensed matter systems e.g. superconductors magnetic field can play a significant role. A striking contrast of the effect of magnetic field 
on the chiral condensate contrary to superconductors is that the magnetic field helps to strengthen the chiral condensate. 
Naively one can 
understand this in the following way. Unlike the electrically charged superconducting condensate, chiral condensate is an electrically
neutral spin zero condensate. Hence for the chiral condensate, the magnetic moment of the fermion and the antifermion
point in the same direction. Hence in the presence of magnetic field both magnetic moments can align themselves along the direction of the 
magnetic field without any frustration in the pair\cite{shovkovynew}. It has also been pointed out that in the presence of magnetic field dimensional reduction 
can play an essential role in the pairing of fermions \cite{dimred1}.

Magnetic catalysis has been explored extensively 
in $(2+1)$ and $(3+1)$- dimensional models with local four fermion interactions \cite{dimred2,dimred3,dimred4,fourfermi1,fourfermi2,fourfermi3,
fourfermi4,fourfermi5,fourfermi6,fourfermi7,fourfermi8,fourfermi9,fourfermi10,fourfermi11,fourfermi12,fourfermi13,fourfermi14,
fourfermi15,fourfermi16,fourfermi17}, supersymmetric models \cite{susy},
quark meson models \cite{pqm1,pqm2}, chiral perturbation theory \cite{chipt2,chipt3} etc.
Such a strong magnetic field can also introduce some exotic phenomenon, e.g. chiral magnetic effect (CME), chiral vortical 
effect (CVE) etc, in a chirally imbalanced medium \cite{kharzeevcme}. Underlying physics of the chiral imbalance is the axial anomaly and 
topologically non trivial vacuum of QCD, which allows topological field configurations like instantons to exist. An asymmetry between the number of left- and right-handed quarks can be generated by these non trivial
topological field configurations due to Adler-Bell-Jackiw (ABJ) anomaly. Such an imbalance can lead to observable $P$ 
and $CP$ violating effects in heavy ion collisions. In the presence  of magnetic field chirally imbalance
quark matter can give rise chiral magnetic effect where a charge separation can be produced.
Effects of a chiral imbalance on the QCD phase diagram can be studied within the framework of 
grand canonical ensemble by introducing a chiral chemical potential $\mu_5$, which enters the QCD Lagrangian via a term 
$\mu_5\bar{\psi}\gamma^0\gamma^5\psi$. Chiral phase transition has been discussed extensively. These studies include 
Nambu-Jona-Lasinio (NJL) type models \cite{chiralNJL1,chiralNJL2,chiralNJL3,chiralNJL4,chiralNJL5,chiralNJL6,chiralNJL7},
quark linear sigma model\cite{chiralNJL1,linearsigma1}, Lattice QCD studies \cite{chiralLattice1,chiralLattice2} etc.
Although the effect of chiral chemical potential has been explored extensively, contradicting results have been reported in 
various literatures, e.g. in Refs.\cite{chiralNJL1,chiralNJL2,chiralNJL3,chiralNJL4,chiralNJL5,chiralNJL6} predict that 
chiral transition temperature decreases with chiral chemical potential. On the other hand in Ref.\cite{chiralNJL7}
it has been argued that with a specific regularization method chiral transition temperature increases with chiral chemical potential,
which is in agreement with Lattice results in Ref.\cite{chiralLattice1,chiralLattice2}. In this context, in a recent interesting work 
the Winger function in presence of non vanishing magnetic field and chiral chemical potential has been evaluated in a non perturbative
manner using explicit solutions of the Dirac equation in a magnetic field and chiral chemical potential \cite{rischke4}. This has been later 
used for pair production in presence of electromagnetic field \cite{shengfang}. 
 
To probe the medium produced in relativistic heavy ion collisions, generally thermodynamic or hydrodynamic model has been used, which
assumes local thermal equilibrium. However, due to the short time scales associated with the strong interaction, 
the medium produced in the heavy ion collision is rather 
dynamical in nature and lives for a very short  time and  
non equilibrium as well as quantum effects can affect the evolution of the medium significantly. 
These effects can be considered within the framework of non equilibrium quantum transport theory. It is important to point out that 
in case of interacting field theory of fermions and gauge bosons, transport theory should be invariant under local gauge transformation. Such a 
gauge covariant quantum transport theory for QCD has been developed in \cite{vasak1,vasak2,kineticqcd1}. 
Classical kinetic theory is characterized by an ensemble of point-like particles with
their single particle phase-space distribution function. The time evolution of single particle phase-space distribution function
governed by the transport equation encodes the evolution of the system. 
Similar to the single particle distribution in classical kinetic theory, Wigner function which is the quantum mechanical analogue of
classical distribution function, encodes quantum corrections in the transport equation \cite{wigner1932}. 
Equation of motion of Winger function can be derived from the equation of motion for the associated field operators, e.g. for fermions, 
evolution equation of Wigner functions can be derived using the Dirac equation\cite{groot,carruthers}. In case of local gauge theories, the 
Wigner function has to be defined in a gauge invariant manner \cite{birula}.
The covariant Wigner function method for spin-1/2 fermions has already been explored extensively in the context of heavy ion collisions
to study various effects including
chiral magnetic effect (CME), chiral vortical effect (CVE), polarization-vorticity coupling, hydrodynamics with spin, dynamical 
generation of magnetic moment etc. \cite{rischke1,lublinsky,rischke2,rischke3,wang,gaowang,teryaev,rischke4,
gorbar2017,shipu,wuhan}.

 In this investigation, we study the chiral phase transition and chiral susceptibility in the presence of magnetic 
 field and chiral chemical potential in quantum kinetic theory framework using Nambu Jona Lasinio (NJL) model
 \cite{njlref1,njlref2,njlref3,njlref4,njlref5,njlref6}.
 Our work is based on the spinor decomposition of the Wigner function using formalism of Ref.\cite{klevanskyneise,rischke4}.
 In this investigation, we limit ourselves to 
 mean field or classical level of the quantum kinetic theory, since the chiral symmetry breaking and generation of dynamical 
 mass of fermions takes place at mean field level \cite{klevanskyneise}. The formulation of transport theory of
 NJL model has been studied in Ref.\cite{klevanskyneise,wilets,abada,florkowski1994}. In this work we have used the 
 formalism given in Ref.\cite{klevanskyneise} to calculate the chiral condensate and the chiral susceptibility
 using the Wigner function. Wigner function in general is used for deriving dynamical equations 
 for the out of equilibrium system \cite{klevanskyneise}. 
 In the present study, we limit ourselves to use the Wigner function for an extended system in global thermal equilibrium i.e. at 
 constant temperature and chemical potentials to calculate chiral susceptibility.
 
 We organize the paper in the following manner. In Sec.\eqref{wignerzeromag}, for the sake of completeness, we recapitulate the results of Ref.\cite{klevanskyneise}
  to study chiral condensate in NJL model using Wigner function approach.
 Then in Sec.\eqref{wignermag} we introduce the Winger 
 function in the presence of magnetic field as well as chiral chemical potential and calculate of the chiral 
 condensate for two flavour NJL model. In Sec.\eqref{chiralsus} we discuss the chiral susceptibility for two flavour NJL model
 in presence of magnetic field ($B$) as well as chiral chemical potential ($\mu_5$).
 In Sec.\eqref{results} we present the results and discussions. Finally in Sec.\eqref{conclu} we conclude our results with an outlook to it.

 \section{Warm up: Wigner function and chiral condensate in NJL model}
 \label{wignerzeromag}
 In this section we first briefly discuss the salient features of the formalism of Wigner function in NJL model for single flavour
 fermion having vanishing current quark mass as given in Ref.\cite{klevanskyneise}.
 Once we get the representation of scalar condensate in terms of Wigner function, we can generalize it to a more realistic situation with 
 non vanishing current quark mass. 
 For a single flavour NJL model we start with the 
 following Lagrangian  \cite{klevanskyneise}, 
 
 \begin{align}
  \mathcal{L}= \bar{\psi}i\slashed{\partial}\psi+G\left((\bar{\psi}\psi)^2+(\bar{\psi} i \gamma_5\psi)^2\right),
 \label{equ1}
  \end{align}
where $\psi$ is the Dirac field, $G$ is the scalar coupling. First terms is the usual kinetic term and the second term represents the four 
Fermi interaction. One can define composite field operators $\hat{\sigma}$ and $\hat{\pi}$ as, 
 \begin{align}
  \hat{\sigma}=-2G\bar{\psi}\psi,~~~~~~~~ \hat{\pi}=-2G\bar{\psi}i\gamma_5\psi.
  \label{equ2}
 \end{align}

 Using Eq.\eqref{equ2}, the Lagrangian given in Eq.\eqref{equ1} can be recasted as  \cite{klevanskyneise}, 
 \begin{align}
  \mathcal{L}=\bar{\psi}i\slashed{\partial}\psi-\hat{\sigma}\bar{\psi}\psi-\hat{\pi}\bar{\psi}i\gamma_5\psi-
  \frac{\hat{\sigma}^2+\hat{\pi}^2}{4G}.
  \label{equ3}
 \end{align}

 In the mean field approximation the operators $\hat{\sigma}$ and $\hat{\pi}$ are replaced by their mean field values, 
 
 \begin{align}
  \hat{\sigma}\rightarrow \sigma = \langle\hat{\sigma}\rangle = \text{Tr}(\hat{\rho}\hat{\sigma}), ~~~
  \hat{\pi}\rightarrow \pi = \langle\hat{\pi}\rangle = \text{Tr}(\hat{\rho}\hat{\pi}),
  \label{equ4}
 \end{align}
where $\hat{\rho}$ is the density matrix operator and ``Tr'' denotes trace over all physical states of the system. 
For a non equilibrium transport theory, in mean field approximation, the fundamental quantity is the Green function, which is defined as,

\begin{align}
 G^{<}_{\alpha\beta}(x,y)=\langle\bar{\psi}_{\beta}(y)\psi_{\alpha}(x)\rangle.
 \label{equ5}
\end{align}
The mean field values of the operators $\hat{\sigma}$ and $\hat{\pi}$, i.e. $\sigma(x)$ and $\pi(x)$ can be determined in terms of the green function 
$ G^{<}(x,y)$ as follows,

\begin{align}
 \sigma(x)=-2G \text{Tr}~ G^{<}(x,x), ~~~~~~~~ \pi(x)=-2G \text{Tr}~ i\gamma_5 G^{<}(x,x).
\label{equ6}
 \end{align}

The Wigner function for fermion is defined as \cite{klevanskyneise},
\begin{align}
  W_{\alpha\beta}(X,p) & =\int\frac{d^4X^{\prime}}{(2\pi)^4}e^{-ip_{\mu}X^{\prime\mu}}\bigg\langle\bar{\psi}_{\beta}\bigg(X+\frac{X^{\prime}}{2}\bigg)
  \psi_{\alpha}\bigg(X-\frac{X^{\prime}}{2}\bigg)\bigg\rangle\nonumber\\
 & = \int\frac{d^4X^{\prime}}{(2\pi)^4}e^{-ip_{\mu}X^{\prime\mu}} G^{<}_{\alpha\beta}\bigg(X+\frac{X^{\prime}}{2},X-\frac{X^{\prime}}{2}\bigg)
 \label{equ7}
 \end{align}

It is important to mention that in NJL model there are no gluons, hence the $SU(3)_c$ gauge invariance
 of the Wigner function does not appear in NJL model. Again in this case we are not considering background magnetic field. 
Hence there is no $U(1)_{em}$ gauge field associated with the NJL model. However in presence of gauge field one has to introduce a 
gauge link in Wigner function for a gauge invariant description \cite{vasakelze}.

Since the Wigner function ($W(X,p)$) as given in Eq.\eqref{equ7}, is a composite operator made out of the Dirac field operators $\psi$ and $\bar{\psi}$, it is convenient to 
decompose $W(X,p)$ in terms of the generators  of the Clifford algebra. The Wigner function $W(X,p)$, in terms of the conventional basis of Clifford
algebra $\mathbb{1},i\gamma_5,\gamma^{\mu},\gamma^{\mu}\gamma_5$ and $\sigma^{\mu\nu}$, can be written as, 

\begin{align}
 W=\frac{1}{4}\bigg[F+i\gamma_5 P+\gamma^{\mu}V_{\mu}+\gamma^{\mu}\gamma^5A_{\mu}+\frac{1}{2}\sigma^{\mu\nu}S_{\mu\nu}\bigg].
 \label{equ8}
\end{align}

Here the coefficients $F, P, V_{\mu}, A_{\mu}$ and $S_{\mu\nu}$ are the  scalar, pseudo scalar, vector, axial vector and tensor 
components of the Wigner function respectively, also known as Dirac-Heisenberg-Wigner (DHW) functions.
The scalar, pseudo scalar, vector, axial vector and tensor Dirac-Heisenberg-Wigner functions can be respectively expressed as,

\begin{align}
 F(X,p)= \text{Tr}~W(X,p),
 \label{equ9}
\end{align}
\begin{align}
 P(X,p)= -i\text{Tr}~\gamma_5W(X,p),
 \label{equ10}
\end{align}
\begin{align}
 V^{\mu}(X,p)= \text{Tr}~\gamma^{\mu}W(X,p),
 \label{equ11}
\end{align}
\begin{align}
 A^{\mu}(X,p)= \text{Tr}~\gamma^5\gamma^{\mu} W(X,p),
 \label{equ12}
\end{align}
\begin{align}
 S^{\mu\nu}(X,p)= \text{Tr}~\sigma^{\mu\nu}W(X,p).
 \label{equ13}
\end{align}

Using Eq.\eqref{equ6}, Eq.\eqref{equ7}, the scalar and pseudo scalar condensates as given in Eq.\eqref{equ9} and Eq.\eqref{equ10}
can be written in terms of Wigner function in the following manner,

\begin{align}
 \sigma(X)=-2G\int d^4p \text{Tr}~W(X,p) = -2G \int d^4p F(X,p),
 \label{equ14}
\end{align}
and,
\begin{align}
 \pi(X)=-2G\int d^4p \text{Tr}~i\gamma_5 W(X,p) = 2G \int d^4p P(X,p).
 \label{equ15}
\end{align}

Using Eq.\eqref{equ2} and Eq.\eqref{equ14} one can express the scalar condensate as,

\begin{align}
 \langle\bar{\psi}\psi\rangle = \int d^4p F(X,p).
 \label{equ16}
\end{align}

In the above description, we have briefly mentioned the relation between the different mean fields with the Wigner function.
It is important to mention that by the virtue of the Dirac equation for the field operator $\psi$ and $\bar{\psi}$ the Wigner function, 
$W(X,p)$, also satisfies a quantum kinetic equation. However in this investigation we have not focused on the kinetic equation of the 
Wigner function. For a detailed discussions on the kinetic equation for the components of Wigner
function, kinetic equation for quark distribution function and related topic see Ref.\cite{klevanskyneise}. In this investigation we rather focus
on the estimation of chiral condensate as given in Eq.\eqref{equ16} and associated chiral 
susceptibility in two flavour NJL model.

The Wigner function can be calculated by inserting the Dirac field operators in Eq.\eqref{equ7}. The Dirac 
field operators in the absence of magnetic field can be written as\cite{Fang2016},
\begin{align}
 \psi(x)=\frac{1}{\sqrt{\Omega}}\sum_{\vec{k},s}\frac{1}{\sqrt{2\mathcal{E}_{0k}}}\left[a(\vec{k},s)u(\vec{k},s)e^{-ik.x}+b^{\dagger}
 (\vec{k},s)v(\vec{k},s)e^{ik.x}\right],
 \label{equ17}
\end{align}
\begin{align}
 \bar{\psi}(x)=\frac{1}{\sqrt{\Omega}}\sum_{\vec{k},s}\frac{1}{\sqrt{2\mathcal{E}_{0k}}}\left[a^{\dagger}(\vec{k},s)\bar{u}(\vec{k},s)e^{ik.x}+b
 (\vec{k},s)\bar{v}(\vec{k},s)e^{-ik.x}\right],
 \label{equ18}
\end{align}
where $\Omega$ is the volume and $s=\pm1$ denotes the spin states. Using the field decomposition as given in Eq.\eqref{equ17} and  Eq.\eqref{equ18},
the Wigner function of a fermion with mass $\mathcal{M}_0$ can be shown to be \cite{Fang2016},

\begin{align}
 W_{\alpha\beta}(X,p)=\frac{1}{(2\pi)^3}\delta(p^2-\mathcal{M}_0^2) & \bigg[\theta(p^0)\sum_sf_{FD}(\mathcal{E}_{0p}-\mu_s)u_{\alpha}(\vec{p},s)\bar{u}_{\beta}(\vec{p},s)\nonumber\\
 & +\theta(-p^0)\sum_s(1-f_{FD}(\mathcal{E}_{0p}+\mu_s))v_{\alpha}(-\vec{p},s)\bar{v}_{\beta}(-\vec{p},s)\bigg],
 \label{equ19}
\end{align}
 where the creation and the annihilation operators of the particle satisfy, $\langle a^{\dagger}(\vec{p},s)a(\vec{p},s)\rangle = 
 f_{FD}(\mathcal{E}_{0p}-\mu_s)$. On the other hand the creation and the annihilation operators of the anti particle satisfy,
 $\langle b^{\dagger}(-\vec{p},s)b(-\vec{p},s)\rangle = 
 f_{FD}(\mathcal{E}_{0p}+\mu_s)$. Here $f_{FD}(z)=1/(1+\exp(z/T))$ is the Fermi Dirac distribution function at temperature $T$ and $\mu_s$ is the chemical potential for the spin state $s$. 
 $\mathcal{E}_{0p}=\sqrt{p^2+\mathcal{M}_0^2}$ is the single particle energy and 
 $\mathcal{M}_0$ is the mass of the Dirac fermion.
  It is important to note that the space time dependence in the Wigner function $W(X,p)$ is hidden in the space time dependence of the temperature
and chemical potential. However for a uniform temperature and chemical potential i.e. for a system in global equilibrium
the Wigner function is independent of space time.
In this investigation we are considering a global thermal equilibrium. Hence from now onward we
will omit the the space time 
dependence in the Wigner function. Using Eq.\eqref{equ9} and Eq.\eqref{equ19} the scalar DHW function can be expressed as \cite{Fang2016}, 

\begin{align}
 F(p)=\mathcal{M}_0\delta(p^2-\mathcal{M}_0^2)\bigg[\frac{2}{(2\pi)^3}\sum_s\bigg(\theta(p^0)f_{FD}(\mathcal{E}_{0p}-\mu_s)
 -\theta(-p^0)(1-f_{FD}(\mathcal{E}_{0p}+\mu_s))\bigg)\bigg].
 \label{equ20}
\end{align}

Using the scalar DHW function as given in Eq.\eqref{equ20}, the scalar condensate for a single fermion species of mass $\mathcal{M}_0$
given in Eq.\eqref{equ16} can be recasted as, 

\begin{align}
 \langle\bar{\psi}\psi\rangle & =  \int d^4p \mathcal{M}_0\delta(p^2-\mathcal{M}_0^2)\bigg[\frac{2}{(2\pi)^3}\sum_s\bigg(\theta(p^0)f_{FD}
 (\mathcal{E}_{0p}-\mu_s)-\theta(-p^0)(1-f_{FD}(\mathcal{E}_{0p}+\mu_s))\bigg)\bigg]\nonumber\\
 & = -\sum_s\int\frac{d^3p}{(2\pi)^3}\frac{\mathcal{M}_0}{\mathcal{E}_{0p}}\bigg[1-f_{FD}(\mathcal{E}_{0p}-\mu_s)
 -f_{FD}(\mathcal{E}_{0p}+\mu_s)\bigg]
 \label{equ21}
\end{align}
In a situation where the chemical potential is independent of the spin of the state, 
\begin{align}
 \langle\bar{\psi}\psi\rangle & = -2N_c\int\frac{d^3p}{(2\pi)^3}\frac{\mathcal{M}_0}{\mathcal{E}_{0p}}
 \bigg[1-f_{FD}(\mathcal{E}_{0p}-\mu)-f_{FD}(\mathcal{E}_{0p}+\mu)\bigg], ~~\text{with}, ~~\mathcal{M}_0=-2G\langle\bar{\psi}\psi\rangle.
 \label{equ23}
\end{align}
The factor of $N_c$ appears in Eq.\eqref{equ23} due to the ``Tr'' over all the degrees of freedom.

Next we shall consider two flavour ($u,d$ quarks) NJL model for vanishing magnetic field and chiral chemical potential, 
with the Lagrangian given as \cite{buballa},

\begin{align}
 \mathcal{L} =  \mathcal{L}_0+ \mathcal{L}_1+ \mathcal{L}_2,
 \label{equ24}
\end{align}
where the free part is, 
\begin{align}
 \mathcal{L}_0 = \bar{\psi}(i\slashed{\partial}-m)\psi,
 \label{equ25}
\end{align}
and the interaction parts are given as,
\begin{align}
 \mathcal{L}_1 = G_1\sum_{a=0}^{3}\bigg[(\bar{\psi}\tau^a\psi)^2
 +(\bar{\psi}i\gamma_5\tau^a\psi)^2\bigg],
 \label{equ26}
\end{align}
and,
\begin{align}
 \mathcal{L}_2 = G_2\left[(\bar{\psi}\psi)^2-(\bar{\psi}\vec{\tau}\psi)^2-(\bar{\psi}i\gamma_5\psi)^2+(\bar{\psi}
 i\gamma_5\vec{\tau}\psi)^2\right],
 \label{equ27}
\end{align}
where $\psi=(\psi_u, \psi_d)^T$ is the quark doublet, $m =\text{diag}(m_u,m_d)$ is the current quark mass with $m_u=m_d$.
$\tau^0=I_{2\times2}$ and $\vec{\tau}$ are the Pauli matrices.
The above Lagrangian as given in Eq.\eqref{equ24} is invariant under $SU(2)_L\times SU(2)_R\times U(1)_V$ transformations. $\mathcal{L}_1$ has an additional $U(1)_A$
symmetry. $\mathcal{L}_2$ is identical with t-Hooft determinant interaction term which breaks the $U(1)_A$ symmetry explicitly. 
$\mathcal{L}_2$ interaction term introduces mixing between different flavours. Value of the coupling $G_2$ is fixed
 by fitting the masses of the pseudoscalar octet \cite{buballa}.
It is also important to emphasis that since we are considering only the scalar 
condensates of the form $\bar{\psi}_u\psi_u$ and $\bar{\psi}_d\psi_d$, so we can safely ignore the pseudo scalar
condensate as well as the scalar condensates of the form $\bar{\psi}_u\psi_d$, $\bar{\psi}_d\psi_u$ etc. 
Using these approximations at the mean field level, the Lagrangian of the two flavour 
NJL model as given in Eq.\eqref{equ24} can be expressed as,

\begin{align}
 \mathcal{L}=\bar{\psi}_u(i\slashed{\partial}-\mathcal{M}_{0_u}) \psi_u+\bar{\psi}_d(i\slashed{\partial}-\mathcal{M}_{0_d})\psi_d
  -2G_1\left(\langle\bar{\psi}_u\psi_u\rangle^2+\langle\bar{\psi}_d\psi_d\rangle^2\right) -
 4G_2\langle\bar{\psi}_u\psi_u\rangle\langle\bar{\psi}_d\psi_d\rangle,
 \label{equ28}
\end{align}
where $u$ and $d$ quark condensates are given as $\langle\bar{\psi}_u\psi_u\rangle$ and $\langle\bar{\psi}_d\psi_d\rangle$ respectively.
The constituent quark masses of 
$u$ and $d$ quarks in terms of the chiral condensates are given as, 

\begin{align}
\mathcal{M}_{0_u} = m_u -4 G_1 \langle\bar{\psi}_u\psi_u\rangle -4 G_2 \langle\bar{\psi}_d\psi_d\rangle,
~~~~\mathcal{M}_{0_d} = m_d -4 G_1 \langle\bar{\psi}_d\psi_d\rangle-4 G_2 \langle\bar{\psi}_u\psi_u\rangle.
\label{equ29}
\end{align} 
 
One can easily generalize the scalar condensate as given in Eq.\eqref{equ23} for single flavour NJL model to  multi flavour NJL model. 
Hence for NJL model of $N_f$ quark flavour and $N_c$ color, the chiral condensate can be written as,

\begin{align}
  \langle\bar{\psi}\psi\rangle_{B=0}^{\mu_5=0} = \sum_{f=1}^{N_f} \langle\bar{\psi_f}\psi_f\rangle_{B=0}^{\mu_5=0}\nonumber
  \end{align}
with,  
  \begin{align}
\langle\bar{\psi_f}\psi_f\rangle_{B=0}^{\mu_5=0}  & = -2N_c\int\frac{d^3p}{(2\pi)^3}\frac{\mathcal{M}_{0_f}}{\mathcal{E}_{0p,f}}
  \bigg[1-f_{FD}(\mathcal{E}_{0p,f}-\mu)-
  f_{FD}(\mathcal{E}_{0p,f}+\mu)\bigg].
 \label{equ30}
  \end{align}
The chiral condensate for $N_f$ flavour NJL model as given in Eq.\eqref{equ30} can also be obtained by first
calculating the thermodynamic potential using the mean field Lagrangian as given in Eq.\eqref{equ28} and then calculating the gap equation using the 
minimization of thermodynamic potential.

\section{Wigner function and chiral condensate in NJL model for non vanishing magnetic field and chiral chemical potential}
\label{wignermag}

In presence of magnetic field ($B$) and chiral chemical potential ($\mu_5$) the Wigner function has been explicitly written down 
in Ref.\cite{rischke4},
using solutions of the Dirac equation for fermions in magnetic field and finite chiral chemical potential.
We shall use them to calculate chiral condensate. For the sake of completeness we write down the 
relevant expressions for the Wigner function.
In the presence of background magnetic field the Wigner function given in Eq.\eqref{equ7} gets modified to a gauge 
invariant Wigner function as \cite{rischke4}, 

\begin{align}
 W_{\alpha\beta}(X,p)=\int\frac{d^4X^{\prime}}{(2\pi)^4} e^{(-ip_{\mu}X^{\prime\mu})}\bigg\langle\bar{\psi}_{\beta}\bigg(X+\frac{X^{\prime}}{2}\bigg)
 U\bigg(A,X+\frac{X^{\prime}}{2},X-\frac{X^{\prime}}{2}\bigg)\psi_{\alpha}\bigg(X-\frac{X^{\prime}}{2}\bigg)\bigg\rangle,
 \label{equ31}
\end{align}
where $U\bigg(A,X+\frac{X^{\prime}}{2},X-\frac{X^{\prime}}{2}\bigg)$ is the gauge link between two space time points 
$\bigg(X-\frac{X^{\prime}}{2}\bigg)$ and $\bigg(X+\frac{X^{\prime}}{2}\bigg)$ for the gauge field $A^{\mu}$. The gauge link has been introduced to make the Wigner function 
gauge invariant. In the presence of homogeneous external magnetic field along the $z$ direction, the gauge link is just a phase. In this case 
the Wigner function simplifies to, 

\begin{align}
 W_{\alpha\beta}(X,p)=\int\frac{d^4X^{\prime}}{(2\pi)^4} e^{(-ip_{\mu}X^{\prime\mu}-iq Byx^{\prime})}\bigg\langle\bar{\psi}_{\beta}\bigg(X+\frac{X^{\prime}}{2}\bigg)
 \otimes\psi_{\alpha}\bigg(X-\frac{X^{\prime}}{2}\bigg)\bigg\rangle,
 \label{equ32}
\end{align}
where $A^{\mu}(X)=(0,-By,0,0)$ is a specific gauge choice of the external magnetic field. $q$ is the charge of the particle and it has been 
taken to be positive. Analogous to the case of vanishing magnetic field Wigner function can be calculated for non vanishing magnetic field 
by using the Dirac field operator in a background magnetic field. The Wigner function in a background magnetic field at finite temperature ($T$),
 chemical potential ($\mu$)and finite chiral chemical potential $(\mu_5)$ has been shown to be \cite{rischke4},

\begin{align}
 W(p)=\sum_{n,s}\left[f_{FD}(E_{p_z,s}^{(n)}-\mu)\delta(p_0+\mu-E_{p_z,s}^{(n)})W_{+,s}^{(n)}(\vec{p})+
 (1-f_{FD}(E_{p_z,s}^{(n)}+\mu))\delta(p_0+\mu+E_{p_z,s}^{(n)})W_{-,s}^{(n)}(\vec{p})\right], ~~~n\geq 0
 \label{equ33}
\end{align}
where the functions $W_{\pm,s}^{(n)}(\vec{p})$ denote the contribution of fermion/anti-fermion in the $n$-th Landau level. The single particle energy 
at the lowest Landau level and higher Landau levels are given as $E_{p_z}^{(0)}= \sqrt{M^2+(p_z-\mu_5)^2}$ and
$E_{p_z,s}^{(n)}=\sqrt{M^2+(\sqrt{p_z^2+2nqB}-s\mu_5)^2}$ respectively. $+$ and $-$ in Eq.\eqref{equ33} denote contributions of positive and negative
energy solutions respectively. In the lowest Landau level fermions can only be in a specific
spin state. On the other hand for higher Landau levels $(n>0)$ both spin states contribute.

The functions $W_{\pm,s}^{(n)}(\vec{p})$ in Eq.\eqref{equ33} can be expressed in terms of Dirac spinors in the following manner\cite{rischke4}, 

\begin{align}
 W_{rs}^{(n)}(\vec{p})\equiv \frac{1}{(2\pi)^3}\int dy^{\prime} \exp(ip_yy^{\prime})\xi_{rs}^{(n)\dagger}
 \left(p_x,p_z,\frac{y^{\prime}}{2}\right)\gamma^0\otimes \xi_{rs}^{(n)}
 \left(p_x,p_z,-\frac{y^{\prime}}{2}\right), ~~~n\geq 0
 \label{equ35}
\end{align}
In Eq.\eqref{equ35}, $r=\pm$ denotes positive energy and negative energy solutions respectively.
The Dirac spinors $\xi^{(0)}_r$ and $\xi^{(n)}_{rs}$, where 
$r=\pm$ denotes positive and negative energy states and $s$ denotes the spin of the state, are defined as,

\begin{align}
 \xi_r^{(0)}(p_x,p_z,y)=\frac{1}{\sqrt{2E_{p_z}^{(0)}}} \begin{pmatrix}
r\sqrt{E_{p_z}^{(0)}-r(p_z-\mu_5)} \\
\sqrt{E_{p_z}^{(0)}+r(p_z-\mu_5)}
\end{pmatrix}
\otimes \chi^{(0)}(p_x,y),
\label{equ36}
\end{align}

\begin{align}
 \xi_{rs}^{(n)}(p_x,p_z,y)=\frac{1}{\sqrt{2E_{p_z,s}^{(n)}}} \begin{pmatrix}
r\sqrt{E_{p_z,s}^{(n)}+r\mu_5-rs\sqrt{p_z^2+2nqB}} \\
\sqrt{E_{p_z,s}^{(n)}-r\mu_5+rs\sqrt{p_z^2+2nqB}}
\end{pmatrix}
\otimes \chi^{(n)}(p_x,p_z,y),~~~~n>0
\label{equ37}
\end{align}
where the normalized eigen spinors $\chi$ are 

\begin{align}
 \chi^{(0)}(p_x,y)= \begin{pmatrix}
                     1 \\
                     0
                    \end{pmatrix}\phi_0 (p_x,y),
\label{equ38}
                    \end{align}
and,
\begin{align}
\chi_s^{(n)}(p_x,p_z,y)=\frac{1}{\sqrt{2\sqrt{p_z^2+2nqB}}}\begin{pmatrix}
                                                            \sqrt{\sqrt{p_z^2+2nqB}+sp_z}\phi_n(p_x,y)\\
                                                            s\sqrt{\sqrt{p_z^2+2nqB}-sp_z}\phi_{n-1}(p_x,y)
                                                           \end{pmatrix}, ~~~~~n>0
                                                    \label{equ39}
\end{align}
where, 

\begin{align}
 \phi_n(p_x,y)=\left(\frac{qB}{\pi}\right)^{1/4}\frac{1}{\sqrt{2^nn!}}\exp\left[-\frac{qB}{2}\left(y+\frac{p_x}{qB}\right)^2\right]
 H_n\left[\sqrt{qB}\left(y+\frac{p_x}{qB}\right)\right], ~~~~n\geq0.
 \label{equ40}
\end{align}
$H_n$ represents $n$-th Hermite polynomial. Inserting the explicit expression of the Dirac spinors as given in Eq.\eqref{equ38} and
Eq.\eqref{equ39} into Eq.\eqref{equ35} one can get the explicit form the function $W_{\pm,s}^{(n)}(\vec{p})$\cite{rischke4}.
For lowest Landau level,

\begin{align}
 W_r^{(0)}(\vec{p})=\frac{r}{4(2\pi)^3E_{p_z}^{(0)}}\Lambda^{(0)}(p_T)\left[M(1+\sigma^{12})+rE_{p_z}^{(0)}(\gamma^0-\gamma^5\gamma^3)
 -(p_z-\mu_5)(\gamma^3-\gamma^5\gamma^0)\right],
 \label{equ41}
\end{align}
while for higher Landau levels, 

\begin{align}
 W_{rs}^{(n)}(\vec{p})=r\frac{1}{4(2\pi)^3E_{p_z,s}^{(n)}} & \bigg\{\bigg[\Lambda_+^{(n)}(p_T)+s\frac{p_z}{\sqrt{p_z^2+2nqB}}\Lambda_-^{(n)}(p_T)
 \bigg]\bigg[M+rE_{p_z,s}^{(n)}\gamma^0+(s\sqrt{p_z^2+2nqB}-\mu_5)\gamma^5\gamma^0\bigg]-\nonumber\\
 &
 \bigg[\Lambda_-^{(n)}(p_T)+s\frac{p_z}{\sqrt{p_z^2+2nqB}}\Lambda_+^{(n)}(p_T)
 \bigg]\bigg[(s\sqrt{p_z^2+2nqB}-\mu_5)\gamma^3+rE_{p_z,s}^{(n)}\gamma^5\gamma^3-M\sigma^{12}\bigg]\nonumber\\
 & -\frac{2nqB}{p_T^2\sqrt{p_z^2+2nqB}}\Lambda_+^{(n)}(p_T)\bigg[(\sqrt{p_z^2+2nqB}-s\mu_5)(p_x\gamma^1+p_y\gamma^2)\nonumber\\
 & 
 +rsE_{p_z,s}^{(n)}(p_x\gamma^5\gamma^1+p_y\gamma^5\gamma^2)-sM(p_z\sigma^{23}-p_y\sigma^{13})\bigg]\bigg\}, ~~~n> 0
 \label{equ42}
\end{align}
where,
\begin{align}
 \Lambda_{\pm}^{(0)}(p_T)=2 \exp\left(-\frac{p_T^2}{qB}\right),
 \label{equ43}
\end{align}
\begin{align}
 \Lambda_{\pm}^{(n)}(p_T)=(-1)^n\left[L_n\left(\frac{2p_T^2}{qB}\right)\mp L_{n-1}\left(\frac{2p_T^2}{qB}\right)\right]
 \exp\left(-\frac{p_T^2}{qB}\right), ~~~n>0.
 \label{equ44}
\end{align}
Here $L_n(x)$ are the Laguerre polynomials with $L_{-1}(x)=0$. Using the Wigner function $W(p)$ as given in Eq.\eqref{equ33}
it can be shown that the scalar DWH function is \cite{rischke4}, 

\begin{align}
F(p)=M\bigg[ \sum_{n=0}^{\infty} V_n(p_0,p_z) \Lambda^{(n)}_{+}(p_T) + \sum_{n=1}^{\infty}
\frac{1}{\sqrt{p_z^2 + 2nqB}} A_n(p_0,p_z) p_z \Lambda^{(n)}_{-}(p_T) \bigg],
\label{equ45}
\end{align}
where, 
\begin{align}
V_0(p_0,p_z) = \frac{2}{(2\pi)^3}\delta\{(p_0+\mu)^2-|E^{(0)}_{p_z}|^2\}\{\theta(p_0+\mu)f_{FD}(p_0) + \theta(-p_0 -\mu)[f_{FD}(-p_0)-1]\}
\label{equ46}
\end{align}

\begin{align}
V_n(p_0,p_z) = \frac{2}{(2\pi)^3}\sum_s\delta\{(p_0+\mu)^2-|E^{(n)}_{p_z,s}|^2\}\{\theta(p_0+\mu)f_{FD}(p_0) +
\theta(-p_0 -\mu)\big[f_{FD}(-p_0)-1\big]\}, ~~~n>0
\label{equ47}
\end{align}

\begin{align}
A_n(p_0,p_z) = \frac{2}{(2\pi)^3}\sum_ss\delta\{(p_0+\mu)^2-|E^{(n)}_{p_z,s}|^2\}\{\theta(p_0+\mu)f_{FD}(p_0) +
\theta(-p_0 -\mu)\big[f_{FD}(-p_0)-1\big]\}, ~~~n>0
\label{equ48}
\end{align}

Once the scalar DWH function is known explicitly as given in Eq.\eqref{equ45}, the chiral condensate of single flavour fermion can be calculated
using Eq.\eqref{equ16} and is given as, 

\begin{align}
 \langle\bar{\psi}\psi\rangle = \int d^4p ~ F(p) = \int 2\pi p_T~dp_0 ~dp_T ~dp_z ~ F(p)
 \label{equ49}
\end{align}

Using Eq.\eqref{equ45} and Eq.\eqref{equ49}, it can be shown that (see Appedix \ref{appendix1} for details), 

\begin{align}
\langle \bar{\psi} \psi \rangle^{\mu_5\neq0}_{B\neq0} & = -\frac{qB}{(2\pi)^2} \bigg[\int dp_z\, \frac{M}{E^{(0)}_{p_z}} \Big[ 1-f_{FD}(E^{(0)}_{p_z}-\mu)\, - 
f_{FD}(E^{(0)}_{p_z}+\mu)\Big]\,\nonumber\\
 & +  \sum_{n=1}^{\infty}\sum_s \int dp_z\,   \frac{M}{E^{(n)}_{p_z,s}}\Big[1-f_{FD}(E^{(n)}_{p_z,s}-\mu) 
 - \,f_{FD}(E^{(n)}_{p_z,s}+\mu)\Big]\bigg]
 \label{equ50}
 \end{align}
 For vanishing chiral chemical potential, $\mu_5=0$, scalar condensate get reduced to,
 \begin{align}
 \langle \bar{\psi} \psi \rangle^{\mu_5=0}_{B\neq0} = -\frac{qB}{(2\pi)^2}  \sum_{n=0}^{\infty} (2-\delta_{n,0})\int dp_z\,   \frac{M_0}
 {\epsilon^{(n)}_{p_z}}\Big[1-f_{FD}(\epsilon^{(n)}_{p_z}-\mu) 
 - \,f_{FD}(\epsilon^{(n)}_{p_z}+\mu)\Big]\bigg],
 \label{equ51}
\end{align}
where, we denote $M_0$ as the mass of fermion in the absence of chiral chemical potential and finite magnitude field. 
The single particle energy $\epsilon^{(n)}_{p_z}$, for vanishing chiral chemical potential can be written as, 

\begin{align}
 \epsilon^{(n)}_{p_z} = \sqrt{M_0^2+p_z^2+2nqB}, ~~n\geq0.
 \label{equ52}
\end{align}

The chiral condensate for a single flavour as given in Eq.\eqref{equ50} can be easily extended to NJL model with two flavours.
Most general Lagrangian for two flavour NJL model with $u$ and $d$ quarks in the magnetic field including
chiral chemical potential is given as, 

\begin{align}
 \mathcal{L}=\bar{\psi}(i\slashed{D}-m+\mu_5\gamma^0\gamma^5)\psi+G_1\sum_{a=0}^3\left[(\bar{\psi}\tau^a\psi)^2+(\bar{\psi}i\gamma_5\tau^a\psi)^2\right]
 +G_2\left[(\bar{\psi}\psi)^2-(\bar{\psi}\vec{\tau}\psi)^2-(\bar{\psi}i\gamma_5\psi)^2+(\bar{\psi}i\gamma_5\vec{\tau}\psi)^2\right],
 \label{equ53}
\end{align}
where $\psi$ is the $U(2)$ quark doublet, given as $\psi=(\psi_u,\psi_d)^T$. The covariant derivative is given as $\slashed{D}=\slashed{\partial}+iq\slashed{A}$ and the current quark mass matrix is $m=\text{diag}(m_u,m_d)$, with $m_u=m_d$. The 
first term in Eq.\eqref{equ53} is the free Dirac Lagrangian in the presence of magnetic field.
For the calculation we have considered the gauge choice of the background magnetic field as $A^{\mu}=(0,-By,0,0)$. 
The second term in Eq.\eqref{equ53} is the four Fermi interaction and the 
attractive part of the quark anti-quark channel of the Fierz transformed color current-current interaction. $\tau^a, a=0,..3$ are the $U(2)$ 
generators in the flavour space. Third term is the t-Hooft interaction terms which introduces flavour mixing as earlier in Eq.\eqref{equ27}.
Since the magnetic field couples to the electric charge of particles, in the presence
of magnetic field $u$ quark and $d$ quarks couple differently with the magnetic field, hence the isospin symmetry 
is explicitly broken. In the mean field approximation, in the absence of any pseudo scalar condensate, Eq.\eqref{equ53}
can be recasted as,

\begin{align}
 \mathcal{L} & =\bar{\psi}_u(i\slashed{D}-M_u+\mu_5\gamma^0\gamma^5) \psi_u+\bar{\psi}_d(i\slashed{D}-M_d+\mu_5\gamma^0\gamma^5)
  \psi_d\nonumber\\
  & ~~~~~~~~~~~~~~~~-2G_1\left(\langle\bar{\psi}_u\psi_u\rangle^2+\langle\bar{\psi}_d\psi_d\rangle^2\right) -
 4G_2\langle\bar{\psi}_u\psi_u\rangle\langle\bar{\psi}_d\psi_d\rangle,
 \label{equ54}
\end{align}
where $u, d $ quark condensates are given as $\langle\bar{\psi}_u\psi_u\rangle$ and $\langle\bar{\psi}_d\psi_d\rangle$ respectively.
The constituent quark masses for 
$u$ and $d$ quarks in terms of the chiral condensates can be given as, 

\begin{align}
M_u = m_u -4 G_1 \langle\bar{\psi}_u\psi_u\rangle-4 G_2 \langle\bar{\psi}_d\psi_d\rangle , 
~~~~M_d = m_d -4 G_1 \langle\bar{\psi}_d\psi_d\rangle-4 G_2 \langle\bar{\psi}_u\psi_u\rangle.
\label{equ55}
\end{align}

Generalizing Eq.\eqref{equ50} for two flavour NJL model, the chiral condensate in the presence of magnetic field and chiral chemical 
potential can be written as, 

\begin{align}
 \langle\bar{\psi}\psi\rangle^{\mu_5\neq0}_{B\neq0}  & = \sum_{f=u,d}\langle\bar{\psi}_f\psi_f\rangle^{\mu_5\neq0}_{B\neq0}, 
 \label{equ56}
\end{align}
where,
\begin{align}
\langle \bar{\psi}_f \psi_f \rangle^{\mu_5\neq0}_{B\neq0} & = -\frac{N_c|q_f|B}{(2\pi)^2} \bigg[\int dp_z\, \frac{M_f}{E^{(0)}_{p_z,f}} 
\Big[ 1-f_{FD}(E^{(0)}_{p_z,f}-\mu)\, - 
f_{FD}(E^{(0)}_{p_z,f}+\mu)\Big]\,\nonumber\\
 & +  \sum_{n=1}^{\infty}\sum_s \int dp_z\,   \frac{M_f}{E^{(n)}_{p_z,s,f}}\Big[1-f_{FD}(E^{(n)}_{p_z,s,f}-\mu) 
 - \,f_{FD}(E^{(n)}_{p_z,s,f}+\mu)\Big]\bigg],
 \label{equ57}
 \end{align}
 and the single particle energy of flavour $f$ can be expressed as,
 
 \begin{align}
E_{p_z,f}^{(0)}= \sqrt{M^2_f+(p_z-\mu_5)^2}~~\text{for}~~n=0, ~~~~E_{p_z,s,f}^{(n)}=\sqrt{M^2_f+(\sqrt{p_z^2+2n|q_f|B}-s\mu_5)^2}~~\text{for}
~~n>0.
\label{equ58}
 \end{align}

 For vanishing chiral chemical potential $\mu_5=0$, the chiral condensate of single flavour can be expressed as, 
 
\begin{align}
 \langle\bar{\psi}_f\psi_f\rangle^{\mu_5=0}_{B\neq0} =  -\frac{N_c|q_f|B}{(2\pi)^2}  \sum_{n=0}^{\infty} (2-\delta_{n,0})\int dp_z\,   
 \frac{M_{0_f}}{\epsilon^{(n)}_{p_z,f}}\Big[1-f_{FD}(\epsilon^{(n)}_{p_z,f}
 -\mu) - \,f_{FD}(\epsilon^{(n)}_{p_z,f}+\mu)\Big]\bigg],
 \label{equ59}
\end{align}
and the single particle energies of flavour $f$ can be expressed as,

\begin{align}
 \epsilon^{(n)}_{p_z,f} =\sqrt{p_z^2+M_{0_f}^2+2n|q_f|B}.
 \label{equ60}
\end{align}

The first term of the quark condensate as given in Eq.\eqref{equ59} contains divergence and needs to be regularized to derive meaningful results.
In usual NJL model at vanishing temperature and chemical potential
such integrals are regularized either by a sharp three momentum cutoff\cite{njlref1,buballa} or a smooth cutoff
\cite{cutoff4,cutoff5,cutoff6}. 
Effective models like NJL model which are non renomalizable have to be complemented with  regularization scheme
with the constraint that the physically meaningful results should be eventually independent of the regularization prescription.
In the presence of magnetic field, continuous momentum dependence 
in two spatial dimensions transverse to the direction of magnetic field, are being replaced by a sum over discretized Landau levels.
Hence a sharp three momentum cutoff in the presence of the magnetic field suffers from cutoff artifact.
Rather a smooth momentum cutoff was used in Ref.\cite{chiralNJL2} in the context of chiral magnetic effects in the PNJL model to avoid such cutoff
artifact. To regularize the vacuum term in Eq.\eqref{equ59}, we follow an elegant procedure that was 
followed in Ref.\cite{regmagfield1} by adding and subtracting a zero magnetic field contribution to the 
chiral condensate which encodes the divergent behaviour. However in such separation the equation satisfies by the constituent
quark mass also occur in the divergent vacuum term. We follow the same procedure here to regulate the divergent
integral arsing in Eq.\eqref{equ59}. The regularized chiral condensate in the presence of magnetic field at vanishing
quark chemical potential is (see Appendix \eqref{appendix2}, Eq.\eqref{appenB16}), 

\begin{align}
 \langle\bar{\psi}_f\psi_f\rangle^{\mu_5=0}_{B\neq0}   = &
 -2N_c\int_{|\vec{p}|\leq\Lambda}\frac{d^3p}{(2\pi)^3}\frac{M_{0_f}}{\sqrt{p^2+M_{0_f}^2}}
 -\frac{N_c M_{0_f} |q_f|B}{2\pi^2}\bigg[x_{0_f}(1-\ln x_{0_f})+\ln \Gamma(x_{0_f})+\frac{1}{2}\ln \bigg(\frac{x_{0_f}}{2\pi}\bigg)\bigg]\nonumber\\
 & +\frac{N_c|q_f|B}{2\pi^2}  \sum_{n=0}^{\infty} (2-\delta_{n,0})\int_{-\infty}^{\infty} dp_z\,   
 \frac{M_{0_f}}{\epsilon^{(n)}_{p_z,f}}f_{FD}(\epsilon^{(n)}_{p_z,f}),
 \label{equ61}
  \end{align}
where the dimensionless variable $x_{0_f}=M_{0_f}^2/2|q_f|B$.  Scalar condensate as given in Eq.\eqref{equ61} can also be obtained by minimizing
the regularized thermodynamic potential using the mean field Lagrangian as given in Eq.\eqref{equ54} in case of vanishing
chiral chemical potential. Solving the equation Eq.\eqref{equ55} using Eq.\eqref{equ61} we get quark masses for vanishing
chiral chemical potential with finite magnetic field. This constituent mass will be later used to estimate 
quark masses at finite chiral chemical potential and finite magnetic field, as discussed in the following subsection.

\subsection{Regularization of chiral condensate in the presence of magnetic field and chiral chemical potential}

Chiral condensate $\langle\bar{\psi}_f\psi_f\rangle$ of quark flavour $f$, in the presence of magnetic field and non zero chiral chemical
potential is given as,

\begin{align}
\langle \bar{\psi}_f \psi_f \rangle^{\mu_5\neq0}_{B\neq0} & = -\frac{N_c|q_f|B}{(2\pi)^2} \bigg[\int dp_z\, \frac{M_f}{E^{(0)}_{p_z,f}} 
\Big[ 1-f_{FD}(E^{(0)}_{p_z,f}-\mu)\, - 
f_{FD}(E^{(0)}_{p_z,f}+\mu)\Big]\,\nonumber\\
 & +  \sum_{n=1}^{\infty}\sum_s \int dp_z\,   \frac{M_f}{E^{(n)}_{p_z,s,f}}\Big[1-f_{FD}(E^{(n)}_{p_z,s,f}-\mu) 
 - \,f_{FD}(E^{(n)}_{p_z,s,f}+\mu)\Big]\bigg]\nonumber\\
 & = \langle \bar{\psi}_f \psi_f \rangle^{\mu_5\neq0}_{vac, B\neq0}+\langle \bar{\psi}_f \psi_f \rangle^{\mu_5\neq0}_{med, B\neq0},
 \label{equ61new}
 \end{align}
 where $\langle \bar{\psi}_f \psi_f \rangle^{\mu_5\neq0}_{vac, B\neq0}$ is zero temperature and zero quark chemical potential part 
 of the chiral condensate and $\langle \bar{\psi}_f \psi_f \rangle^{\mu_5\neq0}_{med, B\neq0}$ is the medium term at finite
 temperature and quark chemical potential. $\langle \bar{\psi}_f \psi_f \rangle^{\mu_5\neq0}_{vac, B\neq0}$  contains 
 divergent integral which has to be regularized to obtain meaningful physical result. To regularize the vacuum part 
 of the chiral condensate for non vanishing magnetic field and chiral chemical potential we have not 
 considered the naive regularization with finite cutoff (Traditional Regularization scheme-TRS) to remove cutoff artifacts,
 rather we have considered Medium Separation Scheme (MSS) outlined in Ref.\cite{regmagfield2}. By adding and subtracting the lowest 
 Landau level term in the zero temperature and zero quark chemical potential 
 part of the chiral condensate for non vanishing magnetic field and chiral chemical potential, we get (for details see
 Appendix \eqref{appendix3}),
  \begin{align}
 \langle \bar{\psi}_f \psi_f \rangle^{\mu_5\neq0}_{vac, B\neq0} & = -\frac{N_c|q_f|B}{(2\pi)^2}\sum_{n=0}^{\infty}\sum_{s=\pm1}\int dp_z 
 \frac{M_f}{E^{(n)}_{p_z,s,f}} +\frac{N_c|q_f|B}{(2\pi)^2} \int dp_z\frac{M_f}{E^{(0)}_{p_z,f}}\nonumber\\
 & = -\frac{N_c|q_f|B}{(2\pi)^2}\sum_{n=0}^{\infty}\sum_{s=\pm1}\int dp_z \left(\frac{1}{\pi}\right)
 \int_{-\infty}^{\infty} dp_4\frac{M_f}{p_4^2+\left(E^{(n)}_{p_z,s,f}\right)^2} +\frac{N_c|q_f|B}{(2\pi)^2} \int dp_z\frac{M_f}{E^{(0)}_{p_z,f}}\nonumber\\
 & = I_1+I_2,
 \label{equ62new}
 \end{align}
where $E^{(n)}_{p_z,s,f}=\sqrt{M_{f}^2+(\sqrt{p_z^2+2n|q_f|B}-s\mu_5)^2}$ and $E^{(0)}_{p_z,f}=\sqrt{M_f^2+(p_z-\mu_5)^2}$.
 Both integrals $I_1$ and $I_2$ are not convergent at large momentum, hence these integrals have to be regularized to 
 get physically meaningful results. In the present investigation we are using medium separation scheme (MSS) to 
 regularize the integrals $I_1$ and $I_2$. MSS method has also been applied for the case of finite chiral chemical potential
 but vanishing magnetic field in Ref.\cite{chiralNJL7}. In the present case we keep both $B\neq0$ and $\mu_5\neq0$ and use the 
 same scheme in the following. Integral $I_1$ can be regularized by adding and subtracting the similar term with 
 magnetic field ($B$) but $\mu_5=0$,
  
 \begin{align}
  \frac{1}{p_4^2+\left(E^{(n)}_{p_z,s,f}\right)^2} & = \frac{1}{p_4^2+\left(\epsilon^{(n)}_{p_z,f}\right)^2}-\frac{1}{p_4^2+\left(\epsilon^{(n)}_{p_z,f}\right)^2}
 + \frac{1}{p_4^2+\left(E^{(n)}_{p_z,s,f}\right)^2}\nonumber\\
 & = \frac{1}{p_4^2+\left(\epsilon^{(n)}_{p_z,f}\right)^2} + \frac{A+2s\mu_5\sqrt{p_z^2+2n|q_f|B}}
 {\Bigg[p_4^2+\left(\epsilon^{(n)}_{p_z,f}\right)^2\Bigg]\Bigg[p_4^2+\left(E^{(n)}_{p_z,s,f}\right)^2\Bigg]},
 \label{equ63new}
 \end{align}
where, $A = M_{0_f}^2-M_f^2-\mu_5^2$ and $\epsilon^{(n)}_{{p_z,f}}=\sqrt{M_{0_f}^2+p_z^2+2n|q_f|B}$. Using the identity given in Eq.\eqref{equ63new} twice we can write the 
integrand of the integral $I_1$, as given in Eq.\eqref{equ62new}, in the following way, 

\begin{align}
 \frac{1}{p_4^2+\left(E^{(n)}_{p_z,s,f}\right)^2} & = \frac{1}{p_4^2+\left(\epsilon^{(n)}_{p_z,f}\right)^2} + 
  \frac{A+2s\mu_5\sqrt{p_z^2+2n|q_f|B}}
 {\Bigg(p_4^2+\left(\epsilon^{(n)}_{p_z,f}\right)^2\Bigg)^2}+ \frac{(A+2s\mu_5\sqrt{p_z^2+2n|q_f|B})^2}
 {\Bigg(p_4^2+\left(\epsilon^{(n)}_{p_z,f}\right)^2\Bigg)^3}\nonumber\\
 & ~~~~~~~~~~+ \frac{(A+2s\mu_5\sqrt{p_z^2+2n|q_f|B})^3}
 {\Bigg(p_4^2+\left(\epsilon^{(n)}_{p_z,f}\right)^2\Bigg)^3\Bigg(p_4^2+\left(E^{(n)}_{p_z,s,f}\right)^2\Bigg)}.
 \label{equ64new}
\end{align}
Performing $p_4$ integration, we 
obtain (for details see Appendix \eqref{appendix2}), 
   \begin{align}
  I_1& = I_{1_{\text{quad}}} -\frac{M_f(M_{0_f}^2-M_f^2+2\mu_5^2)}{2}I_{1_\text{log}} + I_{1_{\text{finite1}}}+I_{1_{\text{finite2}}},
 \label{equ69new}
 \end{align}
where
\begin{align}
 I_{1_{\text{quad}}}  = -\frac{N_c|q_f|B}{(2\pi)^2}\sum_{n=0}^{\infty}\sum_s\int dp_z \frac{M_f}{\epsilon^{(n)}_{p_z,f}},
\end{align}
\begin{align}
 I_{1_\text{log}}= \frac{N_c|q_f|B}{(2\pi)^2}\sum_{n=0}^{\infty}\sum_s \int dp_z \frac{1}{\bigg(\epsilon^{(n)}_{p_z,f}\bigg)^3},
\label{equ70new}
 \end{align}
 \begin{align}
  I_{1_{\text{finite1}}} = -\frac{N_c|q_f|B}{(2\pi)^2}\sum_{n=0}^{\infty}\sum_s \int dp_z \bigg(\frac{3}{8}\bigg)
  \frac{(M_fA^2-4M_f\mu_5^2M_{0_f}^2)}{\bigg(\epsilon^{(n)}_{p_z,f}\bigg)^5 }, 
 \end{align}
and
\begin{align}
 I_{1_{\text{finite2}}} = -\frac{N_c|q_f|B}{(2\pi)^2}\bigg(\frac{15}{16}\bigg)\sum_{n=0}^{\infty}\sum_s \int dp_z
  \int_0^1 dx \frac{(1-x)^2M_f(A+2s\mu_5\sqrt{p_z^2+2n|q_f|B})^3}{\bigg[\left(\epsilon^{(n)}_{p_z,f}\right)^2
 -x(A+2s\mu_5\sqrt{p_z^2+2n|q_f|B})\bigg]^{7/2}}.
\end{align}

The integrals $I_{1_{\text{quad}}}$ and $I_{1_{\text{log}}}$ are divergent at large momentum. On the other hand
$I_{1_{\text{finite1}}}$ and $I_{1_{\text{finite2}}}$ are finite.

In a similar manner the integral $I_2$ in Eq.\eqref{equ62new} we obtain,
\begin{align}
 I_2 & = I_{2_{\text{finite}}}+ I_{2_{\text{log}}},
 \label{equ71new}
 \end{align}
 where,
 \begin{align}
  I_{2_{\text{finite}}} = \left(\frac{1}{2}\right)\frac{N_c|q_f|B}{(2\pi)^2}\int dp_z\int_0^1dx \frac{M_f(A+2p_z\mu_5)}
 {\Bigg[\bigg(\epsilon^{(0)}_{p_z,f}\bigg)^2-x(A+2p_z\mu_5)\Bigg]^{3/2}},
 \end{align}
and,
\begin{align}
 I_{2_{\text{log}}} = \frac{N_c|q_f|B}{(2\pi)^2}\int dp_z\frac{M_f}
 {\epsilon^{(0)}_{p_z,f}}.
\end{align}

Using Eq.\eqref{equ69new} and Eq.\eqref{equ71new}, $\langle \bar{\psi}_f \psi_f \rangle^{\mu_5\neq0}_{vac,B\neq0}$, 
can be expressed as,

\begin{align}
 \langle \bar{\psi}_f \psi_f \rangle^{\mu_5\neq0}_{vac,B\neq0}  & = 
 -\frac{M_f(M_{0_f}^2-M_f^2+2\mu_5^2)}{2}I_{1_\text{log}}+ I_{1_{\text{finite1}}}+I_{1_{\text{finite2}}}+I_{2_{\text{finite}}}
 + I_{\text{quad}},
\end{align}

where,
\begin{align}
 I_{\text{quad}} = I_{1_{\text{quad}}}+I_{2_{\text{log}}}= \frac{M_f}{M_{0_f}}\bigg[-\frac{N_c|q_f|B}{(2\pi)^2}\sum_{n=0}^{\infty}\sum_s\int dp_z 
 \frac{M_{0_f}}{\epsilon^{(n)}_{p_z,f}}+\frac{N_c|q_f|B}{(2\pi)^2}\int dp_z \frac{M_{0_f}}{\epsilon^{(0)}_{p_z,f}}\bigg]
\end{align}

Each integral in $I_{\text{quad}}$ is divergent. Using dimensional regularization it can be regularized to get
(see Appendix \eqref{appendix2}, Eq.\eqref{appenB2} and Eq.\eqref{appenB16} ),

\begin{align}
 I_{\text{quad}} &  = \frac{M_f}{M_{0_f}}\bigg[-\frac{N_c|q_f|B}{(2\pi)^2}\sum_{n=0}^{\infty}\sum_s\int dp_z 
 \frac{M_{0_f}}{\epsilon^{(n)}_{p_z,f}}+\frac{N_c|q_f|B}{(2\pi)^2}\int dp_z \frac{M_{0_f}}{\epsilon^{(0)}_{p_z,f}}\bigg]
 = I_{\text{quad}}^{field}+I_{\text{quad}}^{vac},
\label{equ78new}
 \end{align}

where,
\begin{align}
 I_{\text{quad}}^{field} =
 -\frac{N_cM_f|q_f|B}{2\pi^2}\Bigg[x_{0_f}(1-\ln x_{0_f})+\ln\Gamma(x_{0_f})+\frac{1}{2}\ln\left(\frac{x_{0_f}}{2\pi}\right)\Bigg]
\end{align}
and,
 \begin{align}
 I_{\text{quad}}^{vac} = - \frac{N_cM_f}{2\pi^2}\Bigg[\Lambda\sqrt{M_{0_f}^2+\Lambda^2}-M_{0_f}^2\ln\left(\frac{\Lambda+\sqrt{\Lambda^2+M_{0_f}^2}}
 {M_{0_f}}\right)\Bigg]
 \end{align}
 
Similarly the term $I_{1_\text{log}}$ is divergent at large momentum, hence it has to be regularized. Regularization of $I_{1_\text{log}}$ can 
be done using dimensional regularization. In the dimensional regularization scheme (see Appendix \eqref{appendix2}, 
Eq.\eqref{appenB19}),
\begin{align}
 I_{1_\text{log}} & = \frac{N_c|q_f|B}{(2\pi)^2}\sum_{n=0}^{\infty}\sum_s \int dp_z \frac{1}
 {\bigg(\epsilon^{(n)}_{p_z,f}\bigg)^3}
 = I_{1_\text{log}}^{field}+I_{1_\text{log}}^{vac},
  \label{equ74new}
\end{align}
here,
\begin{align}
 I_{1_\text{log}}^{field} = -\frac{N_c}{2\pi^2}\bigg[-\ln x_{0_f}+\frac{\Gamma^{\prime}(x_{0_f})}{\Gamma(x_{0_f})}\bigg],
\end{align}
and,
\begin{align}
 I_{1_\text{log}}^{vac} = \frac{N_c}{\pi^2}\bigg(\ln\left(\frac{\Lambda}{M_{0_f}}+\sqrt{1+\frac{\Lambda^2}{M_{0_f}^2}}\right)
 -\frac{\Lambda}{\sqrt{\Lambda^2+M_{0_f}^2}}\bigg).
 \end{align}

 Hence the regularized chiral condensate of quark flavour $f$ for finite magnetic field and chiral chemical potential in Medium Separation
Scheme (MSS) for vanishing quark chemical potential can be expressed as, 

\begin{align}
 \langle \bar{\psi}_f\psi_f\rangle^{\mu_5\neq0}_{B\neq0} & = 
 -\frac{M_f(M_{0_f}^2-M_f^2+2\mu_5^2)}{2}I_{1_\text{log}} + I_{1_{\text{finite1}}}+I_{1_{\text{finite2}}}+I_{2_{\text{finite}}}
  + I_{\text{quad}}\nonumber\\
 &+\frac{N_c|q_f|B}{2\pi^2} \bigg[\int_{-\infty}^{\infty} dp_z\, \frac{M_f}{E^{(0)}_{p_z,f}} 
f_{FD}(E^{(0)}_{p_z,f})\, +  \sum_{n=1}^{\infty}\sum_s \int_{-\infty}^{\infty} dp_z\,
\frac{M_f}{E^{(n)}_{p_z,s,f}}f_{FD}(E^{(n)}_{p_z,s,f})\bigg]
 \label{equ76new}
\end{align}
where, regularized $I_{1_\text{log}}$ and $I_{\text{quad}}$ has been given in Eq.\eqref{equ74new} and Eq.\eqref{equ78new} respectively.
This makes the expression for $\langle \bar{\psi}_f\psi_f\rangle^{\mu_5\neq0}_{B\neq0}$ finite which we shall use later for the 
calculation of constituent mass $(M_f)$ for non vanishing magnetic field and chiral chemical potential. Note that for the estimation 
of constituent mass $(M_f)$ for non vanishing magnetic field and chiral chemical potential one requires
constituent mass $M_{0_f}$ for non vanishing magnetic field and vanishing chiral chemical potential, which can be obtained
from Eq.\eqref{equ61}.

\section{chiral susceptibility}
\label{chiralsus}
The fluctuations and the correlations are an important characteristics of any physical system. 
They provide essential information about the effective degrees of freedom and their possible quasi-particle nature.
These fluctuations and correlations are connected with susceptibility. Susceptibility is the response of the system to small external force.
The chiral susceptibility measures the response of the chiral condensate to the infinitesimal change of the current quark mass.
Chiral susceptibility in two flavour NJL model can be defined as, 

\begin{align}
 \chi_c & =\frac{\partial\langle\bar{\psi}\psi\rangle}{\partial m} = \frac{\partial\langle\bar{\psi}_u\psi_u\rangle}{\partial m}
 +\frac{\partial\langle\bar{\psi}_d\psi_d\rangle}{\partial m}= \chi_{cu}+\chi_{cd} 
 \label{equ67}
\end{align}

Using Eq.\eqref{equ55} we get, 

\begin{align}
 \chi_{cu}= \frac{\partial\langle\bar{\psi_u}\psi_u\rangle}{\partial m}= \frac{\partial\langle\bar{\psi_u}\psi_u\rangle}{\partial M_u} 
 \Bigg(1-4G_1 \frac{\partial\langle\bar{\psi_u}\psi_u\rangle}{\partial m} -4 G_2 
 \frac{\partial\langle\bar{\psi_d}\psi_d\rangle}{\partial m}\Bigg),
 \label{equ68}
\end{align}
and, 
\begin{align}
 \chi_{cd}= \frac{\partial\langle\bar{\psi_d}\psi_d\rangle}{\partial m}= \frac{\partial\langle\bar{\psi_d}\psi_d\rangle}{\partial M_d} 
 \Bigg(1-4G_1 \frac{\partial\langle\bar{\psi_d}\psi_d\rangle}{\partial m} -4 G_2 
 \frac{\partial\langle\bar{\psi_u}\psi_u\rangle}{\partial m}\Bigg).
 \label{equ69}
\end{align}
 
Using Eq.\eqref{equ68}, Eq.\eqref{equ69} solving for $\chi_{cu}$ and $\chi_{cd}$, we get, 
\begin{align}
 \chi_{cu} = \frac{\partial\langle\bar{\psi}_u\psi_u\rangle}{\partial M_u} \frac{1-4 G_2 \chi_{cd}}
 {1+4 G_1 \frac{\partial\langle\bar{\psi}_u\psi_u\rangle}{\partial M_u}},
 \label{equ70}
\end{align}
and,
\begin{align}
 \chi_{cd} = \frac{\partial\langle\bar{\psi}_d\psi_d\rangle}{\partial M_d} \frac{1-4 G_2 \chi_{cu}}
 {1+4 G_1 \frac{\partial\langle\bar{\psi}_d\psi_d\rangle}{\partial M_d}}.
 \label{equ71}
\end{align} 

Solving, Eq.\eqref{equ70} and Eq.\eqref{equ71} we get, 

\begin{align}
 \chi_{cu}= \frac{\partial\langle\bar{\psi}_u\psi_u\rangle}{\partial M_u}
 \Bigg(\frac{1+4(G_1-G_2)\frac{\partial\langle\bar{\psi}_d\psi_d\rangle}{\partial M_d}}
 {\left(1+4G_1\frac{\partial\langle\bar{\psi}_u\psi_u\rangle}{\partial M_u}\right)
 \left(1+4G_1\frac{\partial\langle\bar{\psi}_d\psi_d\rangle}{\partial M_d}\right)
 -16 G_2^2\frac{\partial\langle\bar{\psi}_u\psi_u\rangle}{\partial M_u}\frac{\partial\langle\bar{\psi}_d\psi_d\rangle}{\partial M_d}}
 \Bigg),
 \label{equ72}
 \end{align}
\begin{align}
 \chi_{cd} = \frac{\partial\langle\bar{\psi}_d\psi_d\rangle}{\partial M_d}
 \Bigg(\frac{1+4(G_1-G_2)\frac{\partial\langle\bar{\psi}_u\psi_u\rangle}{\partial M_u}}
 {\left(1+4G_1\frac{\partial\langle\bar{\psi}_u\psi_u\rangle}{\partial M_u}\right)
 \left(1+4G_1\frac{\partial\langle\bar{\psi}_d\psi_d\rangle}{\partial M_d}\right)
 -16 G_2^2\frac{\partial\langle\bar{\psi}_u\psi_u\rangle}{\partial M_u}\frac{\partial\langle\bar{\psi}_d\psi_d\rangle}{\partial M_d}}
 \Bigg). 
 \label{equ73}
\end{align}
It is clear from Eq.\eqref{equ72}, Eq.\eqref{equ73} that, to calculate chiral susceptibility for $u$ and $d$ quarks we have to estimate
$\frac{\partial\langle\bar{\psi}_f\psi_f\rangle}{\partial M_{f}}$. However it is important to note that 
like chiral condensate, chiral susceptibility also contains ultraviolet divergence. Hence
$\frac{\partial\langle\bar{\psi}_f\psi_f\rangle}{\partial M_{f}}$ term also has to be regularized to get meaningful results.
Using Eq.\eqref{equ61}, for vanishing chemical potential ($\mu = 0$) and vanishing chiral chemical potential ($\mu_5 = 0$),
in the presence of magnetic field we get, 

\begin{align}
 \frac{\partial\langle\bar{\psi}_f\psi_f\rangle^{\mu_5=0}_{B\neq0}}{\partial M_{0_f}} & = 
 -\frac{2N_c}{(2\pi)^3}\int_{|\vec{p}|\leq\Lambda}d^3p\bigg[\frac{1}{\sqrt{p^2+M_{0_f}^2}}-\frac{M_{0_f}^2}
 {\sqrt{(p^2+M_{0_f}^2)^3}}\bigg]\nonumber\\
 &-\frac{N_c|q_f|B}{2\pi^2}\bigg[x_{0_f}(1-\ln x_{0_f})+\ln\Gamma(x_{0_f})+\frac{1}{2}\ln\bigg(\frac{x_{0_f}}{2\pi}\bigg)\bigg]
 -\frac{N_cM_{0_f}^2}{2\pi^2}\bigg[-\ln x_{0_f}+\frac{1}{2x_{0_f}}+\frac{\Gamma^{\prime}(x_{0_f})}{\Gamma(x_{0_f})}\bigg]\nonumber\\
 & +\sum_{n=0}^{\infty}\frac{N_c|q_f|B}{\pi^2}(2-\delta_{n,0})\int_{0}^{\infty}dp_z
 \bigg[\frac{1}{\epsilon_{p_z,f}^{(n)}}f_{FD}\bigg(\epsilon_{p_z,f}^{(n)}\bigg)-
 \frac{M_{0_f}^2}{(\epsilon_{p_z,f}^{(n)})^3}f_{FD}\bigg(\epsilon_{p_z,f}^{(n)}\bigg)\nonumber\\
 & ~~~~~~~~~~~~~~~~~~~~~~~~~~~~~~~~~~~~~~~~~~~~~~~~~
 -\frac{1}{T}\bigg(\frac{M_{0_f}}{\epsilon_{p_z,f}^{(n)}}\bigg)^2~f_{FD}\bigg(\epsilon_{p_z,f}^{(n)}\bigg)
 \Bigg(1-f_{FD}\bigg(\epsilon_{p_z,f}^{(n)}\bigg)\Bigg)\bigg]\nonumber\\
  \label{equ74}
\end{align}
$ \frac{\partial\langle\bar{\psi}_f\psi_f\rangle^{\mu_5=0}_{B\neq0}}{\partial M_{0_f}}$
as given in Eq.\eqref{equ74} is regularized and it can be used to calculate $\chi_{cu}$, $\chi_{cu}$ and chial susceptibility $\chi_c$
for finite magnetic field, but vanishing chiral chemical potential. To estimate chiral susceptibility at finite magnetic
field as well as non vanishing chiral chemical potential we have to estimate
regularize $\frac{\partial\langle\bar{\psi}_f\psi_f\rangle}{\partial M_{f}}$ at finite $B$ and $\mu_5$. This regularization
has been done using the MSS regularization scheme.

\subsection{Regularization of chiral susceptibility in presence of magnetic field and chiral chemical potential}
For non vanishing magnetic field ($B$) and chiral chemical potential ($\mu_5$) for $\mu=0$, using Eq.\eqref{equ57} the variation 
of chiral condensate with constituent quark mass can be written as,
\begin{align}
 \frac{\partial\langle\bar{\psi}_f\psi_f\rangle^{\mu_5\neq0}_{B\neq0}}{\partial M_f}
 & = \frac{\partial\langle\bar{\psi}_f\psi_f\rangle^{\mu_5\neq 0}_{vac,B\neq 0}}{\partial M_f}
 +\frac{\partial\langle\bar{\psi}_f\psi_f\rangle^{\mu_5\neq 0}_{med,B\neq 0}}{\partial M_f}.
  \label{equ75}
 \end{align}
 Here the first term is the ``vacuum'' term given as,
 \begin{align}
 \frac{\partial\langle\bar{\psi}_f\psi_f\rangle^{\mu_5\neq 0}_{vac,B\neq 0}}{\partial M_f}
 & = -\frac{N_c|q_f|B}{(2\pi)^2}\sum_{n=0}^{\infty}\sum_{s=\pm1}
 \int dp_z \frac{1}{E^{(n)}_{p_z,s,f}}
 + \frac{N_c|q_f|B}{(2\pi)^2} \int dp_z \frac{1}{E^{(0)}_{p_z,f}}\nonumber\\
& +\frac{N_c|q_f|B}{(2\pi)^2}\sum_{n=0}^{\infty}\sum_{s=\pm1}
 \int dp_z \frac{M_f^2}{\bigg(E^{(n)}_{p_z,s,f}\bigg)^{3}}
 - \frac{N_c|q_f|B}{(2\pi)^2} \int dp_z \frac{M_f^2}{\bigg(E^{(0)}_{p_z,f}\bigg)^{3}}\nonumber\\
& = \mathbf{I}_1 + \mathbf{I}_2+\mathbf{I}_3+\mathbf{I}_4,
 \end{align}
 and the medium dependent term is given as,
 \begin{align}
 \frac{\partial\langle\bar{\psi}_f\psi_f\rangle^{\mu_5\neq 0}_{med,B\neq 0}}{\partial M_f} 
 &  = \frac{N_c|q_f|B}{(2\pi)^2}\int dp_z \frac{1}{E_{p_z,f}^{(0)}}\left(2f_{FD}(E_{p_z,f}^{(0)})\right)
 -\frac{N_c|q_f|B}{(2\pi)^2}\int dp_z \frac{M_{f}^2}{(E_{p_z,f}^{(0)})^3}\left(2f_{FD}(E_{p_z,f}^{(0)})\right)\nonumber\\
 & -\frac{N_c|q_f|B}{(2\pi)^2}\int dp_z \frac{M_{f}^2}{(E_{p_z,f}^{(0)})^2}\left(\frac{2}{T}\right)f_{FD}(E_{p_z,f}^{(0)})
 \left(1-f_{FD}(E_{p_z,f}^{(0)})\right)\nonumber\\
 & +\frac{N_c|q_f|B}{(2\pi)^2}\sum_{n=1}^{\infty}\sum_{s=\pm1}\int dp_z \frac{1}{E_{p_z,s,f}^{(n)}}
 \left(2f_{FD}(E_{p_z,s,f}^{(n)})\right)
 -\frac{N_c|q_f|B}{(2\pi)^2}\sum_{n=1}^{\infty}\sum_{s=\pm1}\int dp_z \frac{M_{f}^2}{(E_{p_z,s,f}^{(n)})^3}\left(2f_{FD}(E_{p_z,s,f}^{(n)})\right)\nonumber\\
 & -\frac{N_c|q_f|B}{(2\pi)^2}\sum_{n=1}^{\infty}\sum_{s=\pm1}\int dp_z \frac{M_{f}^2}{(E_{p_z,s,f}^{(n)})^2}\left(\frac{2}{T}\right)f_{FD}(E_{p_z,s,f}^{(n)})
 \left(1-f_{FD}(E_{p_z,s,f}^{(n)})\right)
 \label{equ92new}
 \end{align}
 The medium dependent term is convergent and does not need any regularization. The ``vacuum'' term on the other hand the integrals,
 $\mathbf{I}_1, \mathbf{I}_2$, and $\mathbf{I}_3$ are are divergent and need regularization. We perform the MSS scheme as was done for the 
 chiral condensate. The regularized $\frac{\partial\langle\bar{\psi}_f\psi_f\rangle^{\mu_5\neq0}_{vac,B\neq0}}{\partial M_f}$ can be
 expressed as (see Appendix \eqref{appendix4}, Eq.\eqref{appenD21}),
 
 \begin{align}
 \frac{\partial\langle\bar{\psi}_f\psi_f\rangle^{\mu_5\neq0}_{vac,B\neq0}}{\partial M_f} 
& = -\left(\frac{M_{0_f}^2-M_f^2+2\mu_5^2}{2}\right)\mathbf{I}_{1,\text{log}} +\mathbf{I}_{1,\text{finite1}}+\mathbf{I}_{1,\text{finite2}} 
 +\mathbf{I}_{2,\text{finite}}+\mathbf{I}_{3,\text{finite}}\nonumber\\
&~~~~~~~~~~~~ + \mathbf{I}_{\text{finite}}
+ \mathbf{I}_{\text{quad}}+\mathbf{I}_{\text{log}},
\label{equ93new}
 \end{align}
where regularized $\mathbf{I}_{\text{quad}}, \mathbf{I}_{\text{log}}, \mathbf{I}_{1,\text{log}}$
can be expressed as (see Appendix \eqref{appendix4}, Eq.\eqref{appenD25}, Eq.\eqref{appenD26}, Eq.\eqref{appenD27}), 
\begin{align}
 \mathbf{I}_{\text{quad}} &  = -\frac{N_c|q_f|B}{(2\pi)^2}\sum_{n=0}^{\infty}\sum_{s=\pm1}\int dp_z \frac{1}{\epsilon^{(n)}_{p_z,f}}
 + \frac{N_c|q_f|B}{(2\pi)^2}\int dp_z \frac{1}{\epsilon^{(0)}_{p_z,f}}\nonumber\\
 & = -\frac{N_c|q_f|B}{2\pi^2}\bigg[x_{0_f}(1-\ln x_{0_f})+\ln\Gamma(x_{0_f})+\frac{1}{2}\ln \left(\frac{x_{0_f}}{2\pi}\right)\bigg]
 -\frac{N_c}{2\pi^2}\bigg[\Lambda\sqrt{\Lambda^2+M_{0_f}^2}-M_{0_f}^2\ln\left(\frac{\Lambda+\sqrt{\Lambda^2+M_{0_f}}}{M_{0_f}}\right)\bigg],
\end{align}
\begin{align}
 \mathbf{I}_{\text{log}} & = \frac{N_c|q_f|B}{(2\pi)^2}\sum_{n=0}^{\infty}\sum_{s=\pm1}\int dp_z \frac{M_f^2}{(\epsilon^{(n)}_{p_z,f})^3}
 - \frac{N_c|q_f|B}{(2\pi)^2}\int dp_z \frac{M_f^2}{(\epsilon^{(0)}_{p_z,f})^3}\nonumber\\
 & = -\frac{N_cM_f^2}{2\pi^2}\bigg[-\ln x_{0_f}+\frac{1}{2x_{0_f}}+\frac{\Gamma^{\prime}(x_{0_f})}{\Gamma(x_{0_f})}\bigg]
 +\frac{N_cM_f^2}{\pi^2}\bigg[\ln\left(\frac{\Lambda+\sqrt{\Lambda^2+M_{0_f}^2}}{M_{0_f}}\right)-\frac{\Lambda}
 {\sqrt{\Lambda^2+M_{0_f}^2}}\bigg],
\end{align}
\begin{align}
 \mathbf{I}_{1,\text{log}} & = \frac{N_c|q_f|B}{(2\pi)^2}\sum_{n=0}^{\infty}\sum_{s=\pm1} \int dp_z \frac{1}{(\epsilon^{(n)}_{p_z,f})^3}\nonumber\\
 & = -\frac{N_c}{2\pi^2}\bigg[-\ln x_{0_f}+\frac{\Gamma^{\prime}(x_{0_f})}{\Gamma(x_{0_f})}\bigg]+
 \frac{N_c}{\pi^2}\bigg[\ln\left(\frac{\Lambda+\sqrt{\Lambda^2+M_{0_f}^2}}{M_{0_f}}\right)-\frac{\Lambda}
 {\sqrt{\Lambda^2+M_{0_f}^2}}\bigg],
 \end{align}
and the convergent integrals $\mathbf{I}_{1,\text{finite1}}$, $\mathbf{I}_{1,\text{finite2}}$,$\mathbf{I}_{2,\text{finite}}$,
$\mathbf{I}_{3,\text{finite}}$ and $\mathbf{I}_{\text{finite}}$ are given as,
\begin{align}
 \mathbf{I}_{1,\text{finite1}} = -\frac{N_c|q_f|B}{(2\pi)^2}\sum_{n=0}^{\infty}\sum_{s=\pm1}\int dp_z \left(\frac{3}{8}\right)\frac{A^2-4\mu_5^2M_f^2}
 {\left(\epsilon^{(n)}_{p_z,f}\right)^5},
\end{align}
\begin{align}
 \mathbf{I}_{1,\text{finite2}} =  -\frac{N_c|q_f|B}{(2\pi)^2}\left(\frac{15}{16}\right)\sum_{n=0}^{\infty}\sum_{s=\pm1}\int dp_z\int_0^1dx 
 \frac{(1-x)^2(A+2s\mu_5\sqrt{p_z^2+2n|q_f|B})^3}{\Bigg[(\epsilon^{(n)}_{p_z,f})^2-x(A+2s\mu_5\sqrt{p_z^2+2n|q_f|B})\Bigg]^{7/2}}
\end{align}
\begin{align}
 \mathbf{I}_{2,\text{finite}} = \left(\frac{1}{2}\right) \frac{N_c|q_f|B}{(2\pi)^2}\int dp_z\int_0^1 dx \frac{A+2p_z\mu_5}{\bigg[(\epsilon^{(0)}_{p_z,f})^2
 -x(A+2p_z\mu_5)\bigg]^{3/2}},
\end{align}
\begin{align}
 \mathbf{I}_{3,\text{finite}} = \frac{N_c|q_f|B}{(2\pi)^2}\sum_{n=0}^{\infty}\sum_{s=\pm1}\int dp_z M_f^2 
 \bigg(\frac{1}{(E^{(n)}_{p_z,s,f})^3}-\frac{1}{(\epsilon^{(n)}_{p_z,f})^3}\bigg),
\end{align}
\begin{align}
 \mathbf{I}_{\text{finite}}
 = -\frac{N_c|q_f|B}{(2\pi)^2}\int dp_z M_f^2 \bigg(\frac{1}{(E^{(0)}_{p_z,f})^3}-\frac{1}{(\epsilon^{(0)}_{p_z,f})^3}\bigg).
\end{align}

For non vanishing magnetic field and chiral chemical potential Eq.\eqref{equ92new}, Eq.\eqref{equ93new} along with 
Eq.\eqref{equ72} and Eq.\eqref{equ73} can be used to calculate chiral susceptibility ($\chi_c$).

\section{results}
\label{results}
Let us note that the Lagrangian as given in Eq.\eqref{equ53} has the following parameters, two couplings $G_1$, $G_2$, the three momentum cutoff 
$\Lambda$ and the current quark masses $m_u$ and $m_d$. For numerical evaluations the two couplings $G_1$, $G_2$, are parametrized 
as $G_1= (1-\alpha)g$ and $G_2= \alpha g$ \cite{buballa,frankbuballa2003}. 
The parameter $\alpha$ is the measure of the strength of the instanton interaction. The other parameters on 
which the chiral condensate depends are, the current quark masses: $m_u=m_d = 6$ MeV, the three momentum cut off : $\Lambda=590$MeV
and the scalar coupling: $g = 2.435/\Lambda^2$. For these values of the parameters, pion vacuum mass is 140.2 MeV, pion decay
constant is 92.6 MeV and the quark condensates are
$\langle\bar{\psi}_u\psi_u\rangle=\langle\bar{\psi}_d\psi_d\rangle$ = $(-241.5)$MeV$^3$.
 This parameter set also leads to a vacuum constituent quark mass $400$ MeV. It is important to mention that in the 
absence of magnetic field $\langle\bar{\psi}_u\psi_u\rangle$ and $\langle\bar{\psi}_d\psi_d\rangle$ are same. Only in the presence of 
magnetic field $\langle\bar{\psi}_u\psi_u\rangle$ and $\langle\bar{\psi}_d\psi_d\rangle$ are different due to the different electric 
charge of $u$ and $d$ quark. Hence in the absence of magnetic field, the constituent quark masses $M_u$ and $M_d$ are same and the constituent mass
of quarks does not depend on the parameter $\alpha$. However $\eta-\eta^{\prime}$ splitting indicates $\alpha = 0.15$. 
Since the quark masses and the associated susceptibilities depend upon $\alpha$ for non vanishing magnetic field, we have taken 
$\alpha$ as a parameter and have shown the results for different values of $\alpha$ to highlight the effect of flavour mixing
interaction.

\begin{figure}[!htb]
    \centering
    \begin{minipage}{.5\textwidth}
        \centering
        \includegraphics[width=1.1\linewidth]{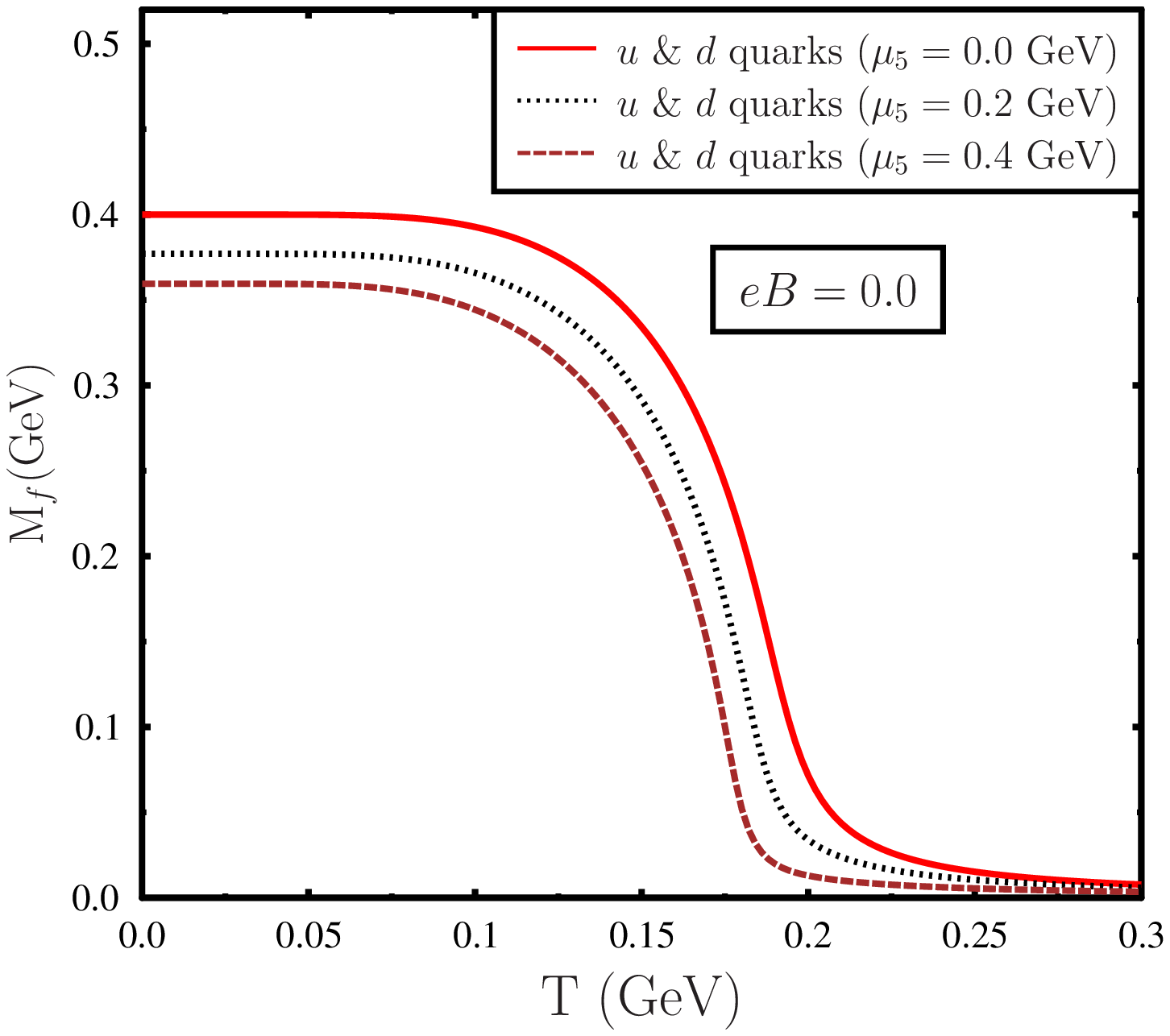}
    \end{minipage}%
    \begin{minipage}{0.5\textwidth}
        \centering
        \includegraphics[width=1.1\linewidth]{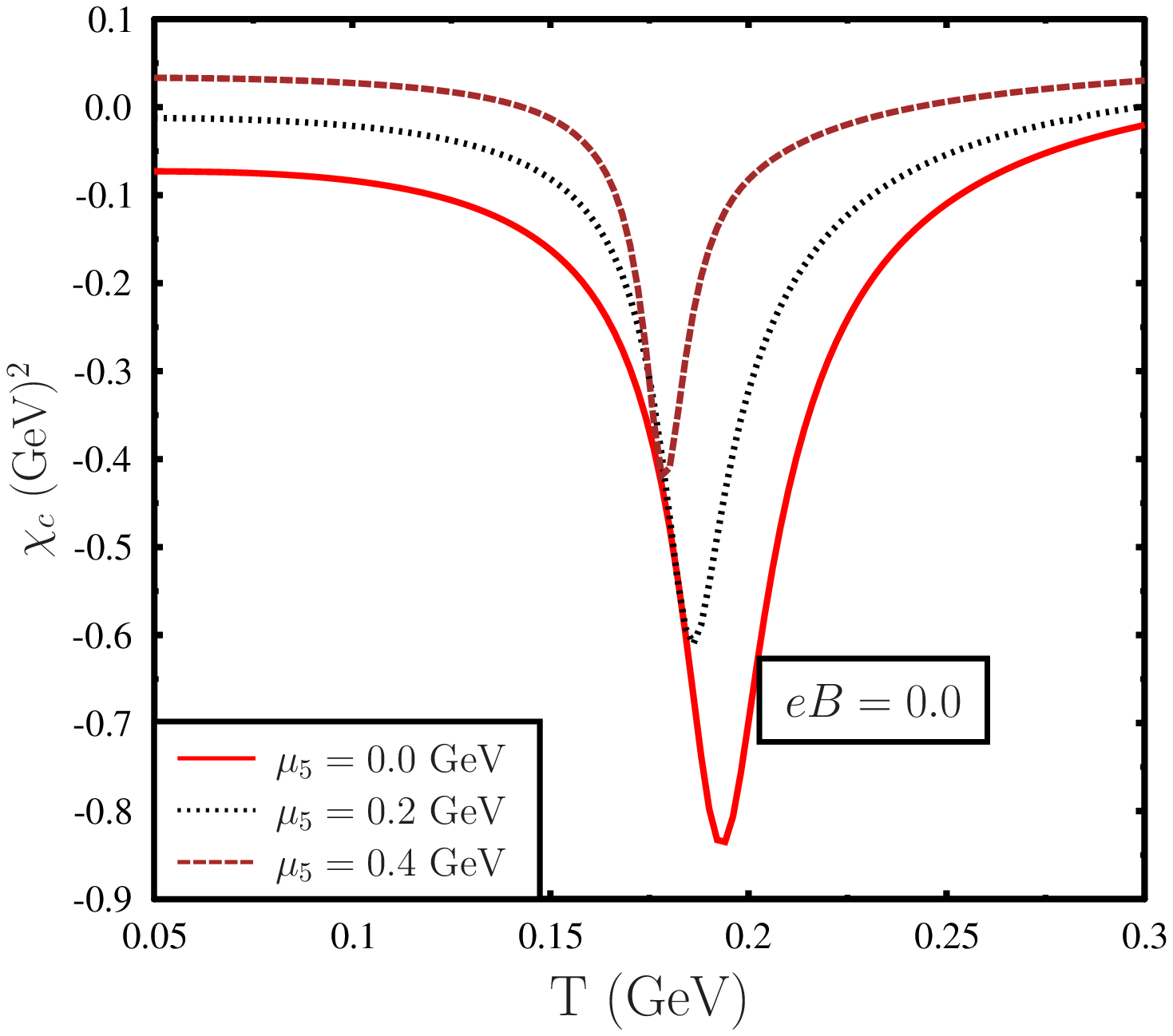}
    \end{minipage}
    \caption{Left plot: Variation of constituent quark mass $M_u=M_d$ with temperature ($T$) for
    zero magnetic field but for various values of chiral chemical potential. 
    Right plot: Variation of chiral susceptibility $\chi_c$ with temperature ($T$)
    for zero magnetic field but with different values of chiral chemical potential. Prominent peak in the 
chiral susceptibility plot shows the chiral transition temperature. From the left plot it is clear that with increasing 
chiral chemical potential ($\mu_5$) constituent mass decreases. From the susceptibility plot it is clear that 
transition temperature decreases with chiral chemical potential.}
    \label{fig2}
\end{figure}

In Fig.\eqref{fig2} we show the variation of constituent quark masses and the associated chiral 
susceptibility as a function of temperature ($T$) for different values of chiral chemical potential ($\mu_5$)
and for vanishing magnetic field.
For zero magnetic field  
$\langle\bar{\psi}_u\psi_u\rangle=\langle\bar{\psi}_d\psi_d\rangle$, hence the masses of the $u$ and $d$ quarks remain same.
From the 
left plot in Fig.\eqref{fig2} we can see that the constituent mass decreases with increasing chiral chemical potential.
One can understand this result in the following way, chiral chemical potential is a conserved number of the associated
chiral symmetry. Chiral symmetry is exact when the fermions are massless. Chiral symmetry try to protect the mass
of the fermion from quantum corrections. Hence the chiral chemical potential which is the measure of the chiral symmetry tries to reduce the 
mass of the fermion. Right plot in Fig.\eqref{fig2} shows the chiral susceptibility for vanishing quark chemical potential 
and magnetic field. Peak in the chiral susceptibility plot shows the chiral transition temperature. Using Eq.\eqref{equ72} and 
Eq.\eqref{equ73} it can be shown that $\chi_{cu} = \chi_{cd}$ for  vanishing magnetic field. Hence 
the variation of total chiral susceptibility ($\chi_{c}$) with temperature shows only one peak. However in the 
presence of magnetic field in general $\chi_{cu}$ can be different from $\chi_{cd}$ and variation
of total chiral susceptibility $\chi_c$ with temperature can show multiple peaks. 
Results for non vanishing magnetic field will be shown later. From the right plot in 
Fig.\eqref{fig2} it is clear that with increasing chiral chemical potential ($\mu_5$) chiral transition temperature decreases.

\begin{figure}[!htp]
\begin{center}
\includegraphics[width=0.75\textwidth]{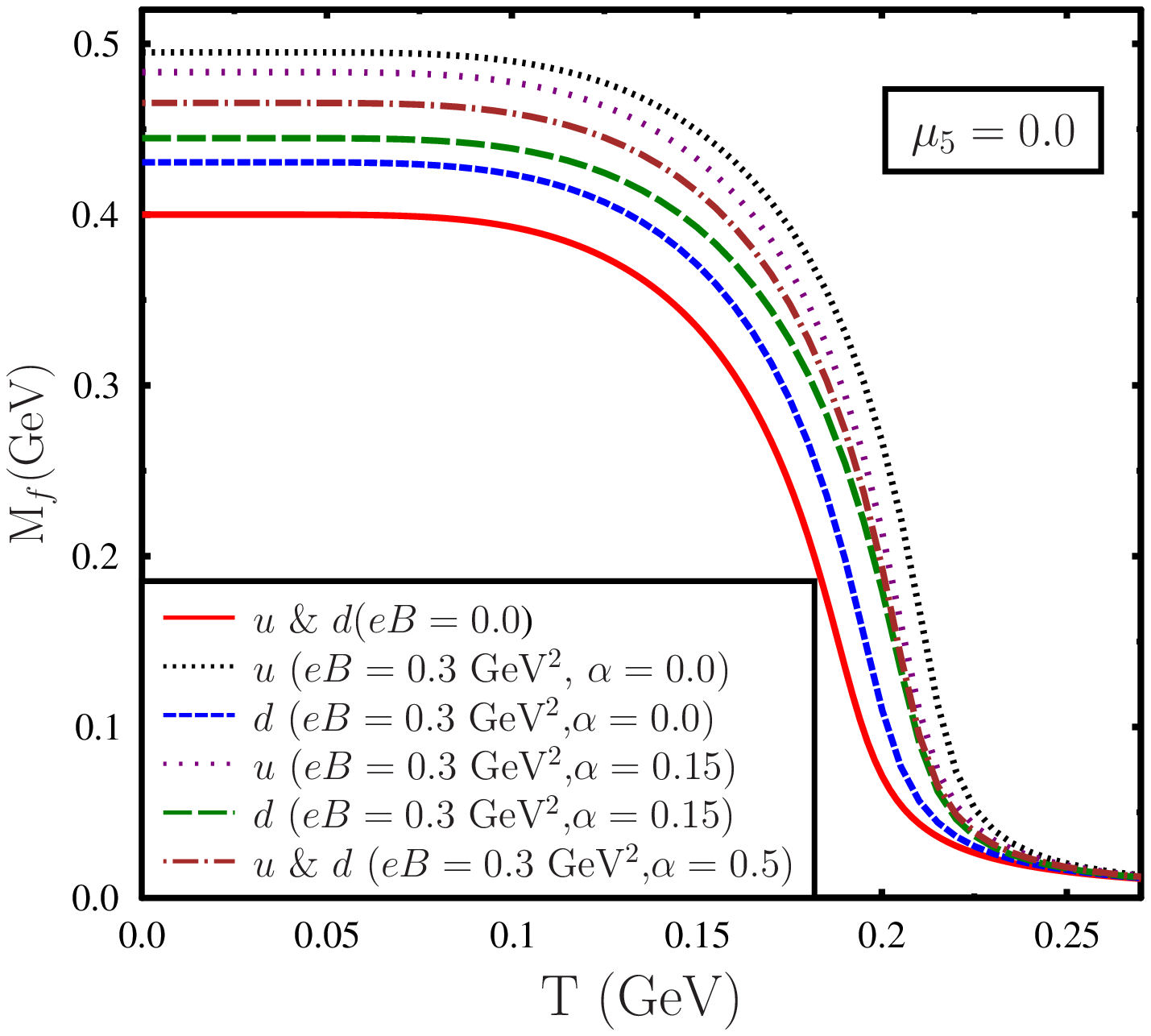}
\caption{Variation of constituent quark masses $M_u$ and $M_d$ with temperature for vanishing
chiral chemical potential but with finite magnetic field for different values of $\alpha$. 
For vanishing magnetic field $M_u$ and $M_d$ are same. Note that in the presence of magnetic field, for $\alpha = 0.5$, although
$\langle\bar{\psi}_u\psi_u\rangle\neq \langle\bar{\psi}_d\psi_d\rangle$, but the constituent quark masses $M_u=M_d$. However 
for $\alpha\neq0.5$, the constituent quark masses $M_u\neq M_d$ in the presence of magnetic field.$\alpha = 0.0$ corresponds to the case when there is no flavour mixing interaction,
and $\alpha=0.5$ corresponds to maximal flavour mixing.} 
\label{fig3}
 \end{center}
 \end{figure}

In Fig.\eqref{fig3} we show the variation of constituent quark masses $M_u$ and $M_d$ with temperature for  vanishing
chiral chemical potential and with finite magnetic field for different values of $\alpha$. From this figure it is clear that 
at non vanishing magnetic field constituent quark mass increases. At vanishing magnetic field constituent mass 
of $u$ and $d$ quarks are same. Although in the presence of magnetic field, quark condensates
$\langle\bar{\psi}_u\psi_u\rangle\neq \langle\bar{\psi}_d\psi_d\rangle$, but for $\alpha=0.5$ the quark masses $M_u=M_d$. This is because
for $\alpha=0.5$ constituent quark mass is $M_f=m-2g(\langle\bar{\psi}_u\psi_u\rangle +
\langle\bar{\psi}_d\psi_d\rangle)$, as can be seen from Eq.\eqref{equ55}. On the other hand for $\alpha\neq0.5$ quark masses $M_u$
and $M_d$ are not the same. The difference between $M_u$ and $M_d$ increases with decrease in the value of $\alpha$ and this 
difference is largest when $\alpha = 0.0$. $\alpha = 0.0$ corresponds to the case when there is no flavour mixing interaction,
and $\alpha=0.5$ corresponds to maximal flavour mixing. It is important to note that for vanishing magnetic field 
flavour mixing interaction does not affect the quark masses. Only in the presence of magnetic field when 
$\langle\bar{\psi}_u\psi_u\rangle\neq \langle\bar{\psi}_d\psi_d\rangle$, flavour mixing interaction affects the 
constituent quark masses $M_u$ and $M_d$ significantly.

\begin{figure}[!htb]
    \centering
    \begin{minipage}{.5\textwidth}
        \centering
        \includegraphics[width=1.1\linewidth]{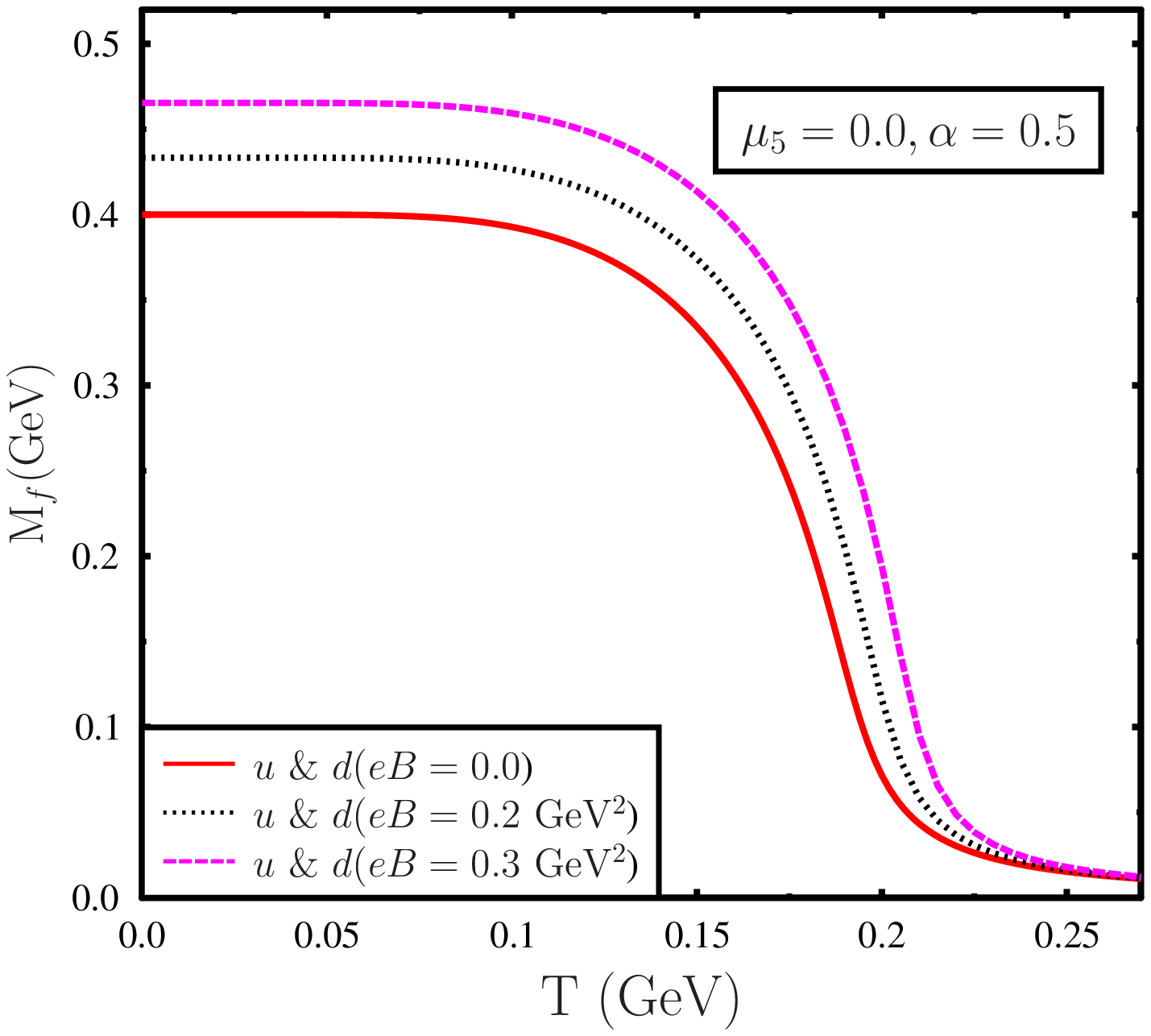}
    \end{minipage}%
    \begin{minipage}{0.5\textwidth}
        \centering
        \includegraphics[width=1.1\linewidth]{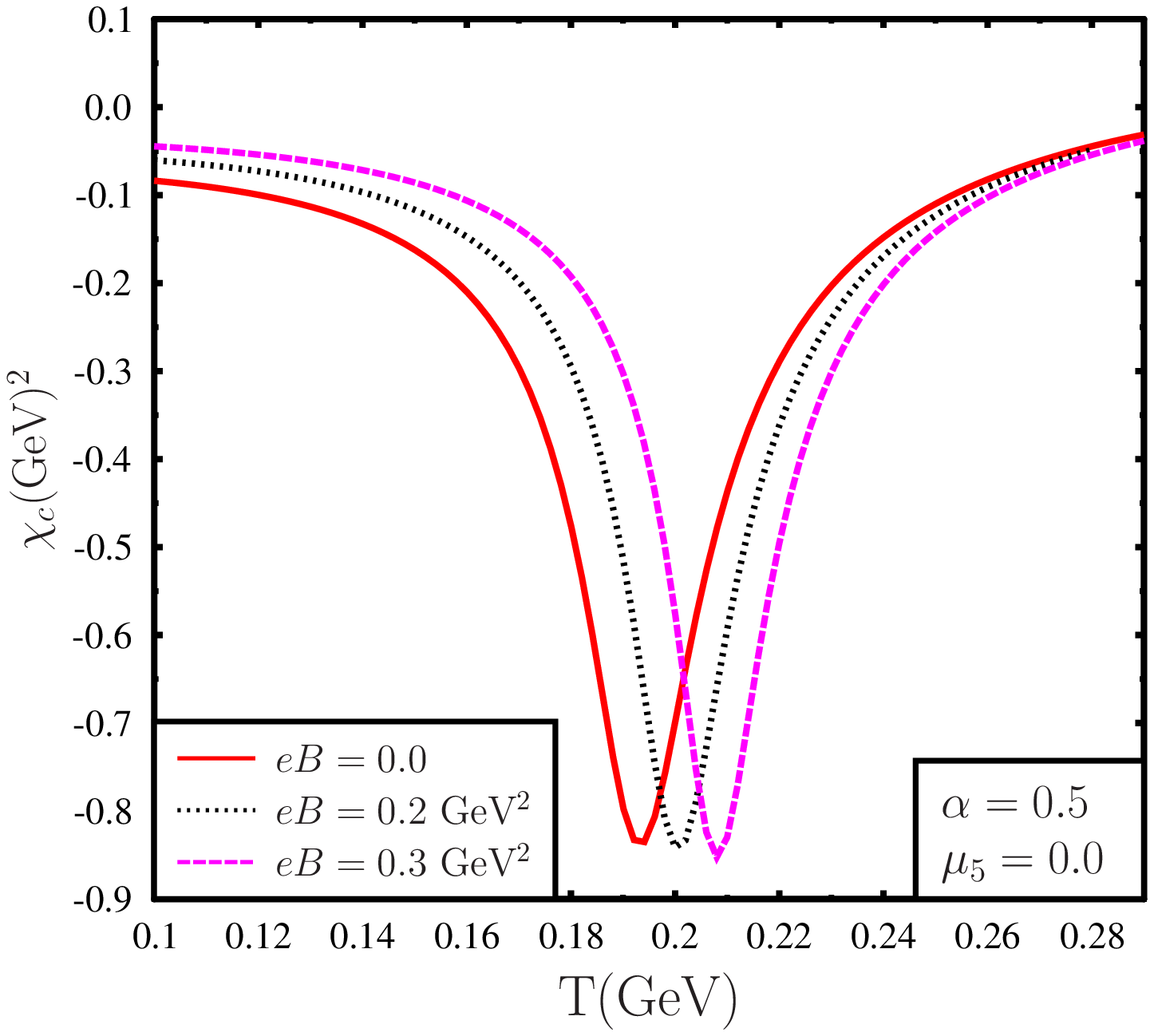}
    \end{minipage}
    \caption{Left plot: variation of constituent quark mass $M_u$ and $M_d$,
with temperature for vanishing chiral chemical potential,
but with different values of magnetic field for $\alpha=0.5$. Right plot: Variation of chiral susceptibility
$\chi_c$ with temperature ($T$)  for vanishing chiral chemical potential,
but with different values of magnetic field for $\alpha=0.5$. From the left plot it is clear that with increasing 
magnetic field constituent mass $M_u=M_d$ increases. From the susceptibility plot it is clear that 
transition temperature increases with magnetic field.}
    \label{fig4}
\end{figure}

In Fig.\eqref{fig4} we show the variation of constituent quark masses $M_u$ and $M_d$ and the associated total chiral susceptibility,
with temperature for vanishing chiral chemical potential and with different values of magnetic field for $\alpha=0.5$. It has been already mentioned that for $\alpha = 0.5$ even in 
the presence of magnetic field $M_u=M_d$. From the left plot in Fig.\eqref{fig4} it is clear that with increasing magnetic field 
constituent quark mass increases. On the other hand from the right plot in Fig.\eqref{fig4} it is clear that chiral transition 
temperature increases with increasing magnetic field.

\begin{figure}[!htb]
    \centering
    \begin{minipage}{.5\textwidth}
        \centering
        \includegraphics[width=1.1\linewidth]{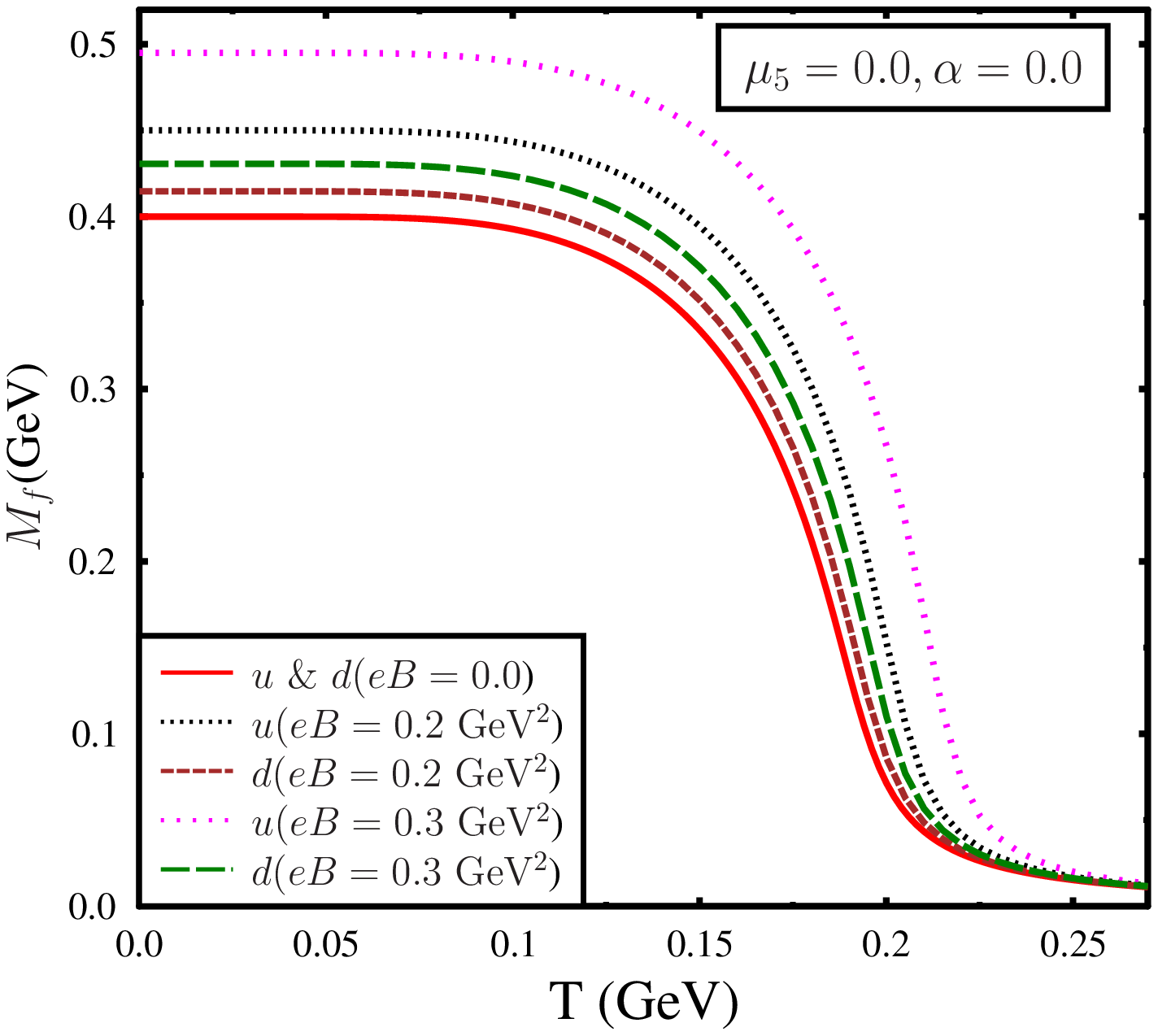}
    \end{minipage}%
    \begin{minipage}{0.5\textwidth}
        \centering
        \includegraphics[width=1.1\linewidth]{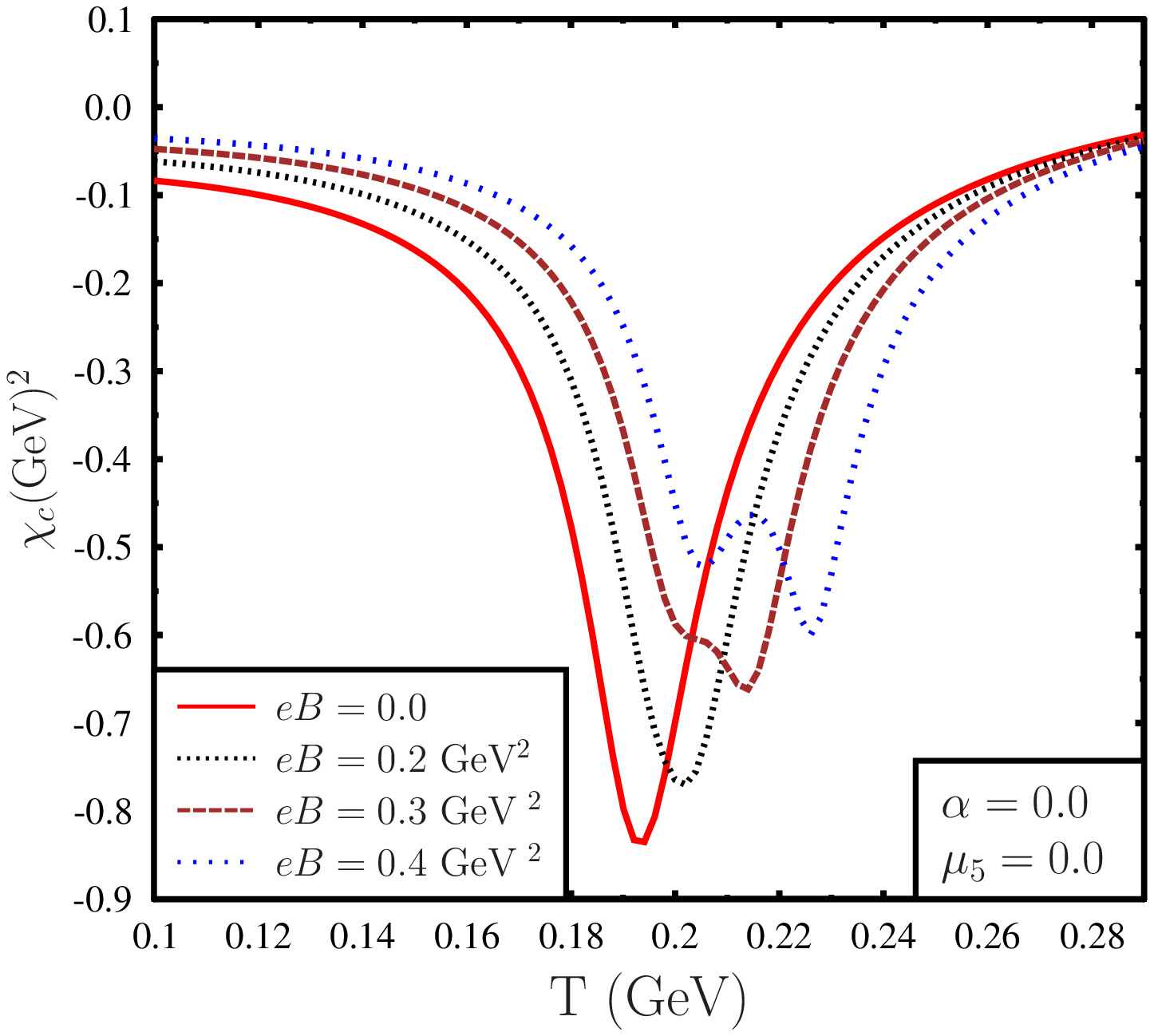}
    \end{minipage}
    \caption{Left plot: variation of constituent quark mass $M_u$ and $M_d$,
with temperature for vanishing chiral chemical potential,
but with different values of magnetic field for $\alpha=0.0$. Right plot: Variation of chiral susceptibility
$\chi_c$ with temperature ($T$)  for vanishing chiral chemical potential,
but with different values of magnetic field for $\alpha=0.5$. From the left plot it is clear that with increasing 
magnetic field constituent mass increases. From the susceptibility plot it is clear that 
transition temperature increases with magnetic field. In the right plot we can observe two distinct peaks at 
relatively large magnetic fields.}
    \label{fig5}
\end{figure}

\begin{figure}[!htb]
    \centering
    \begin{minipage}{.5\textwidth}
        \centering
        \includegraphics[width=1.1\linewidth]{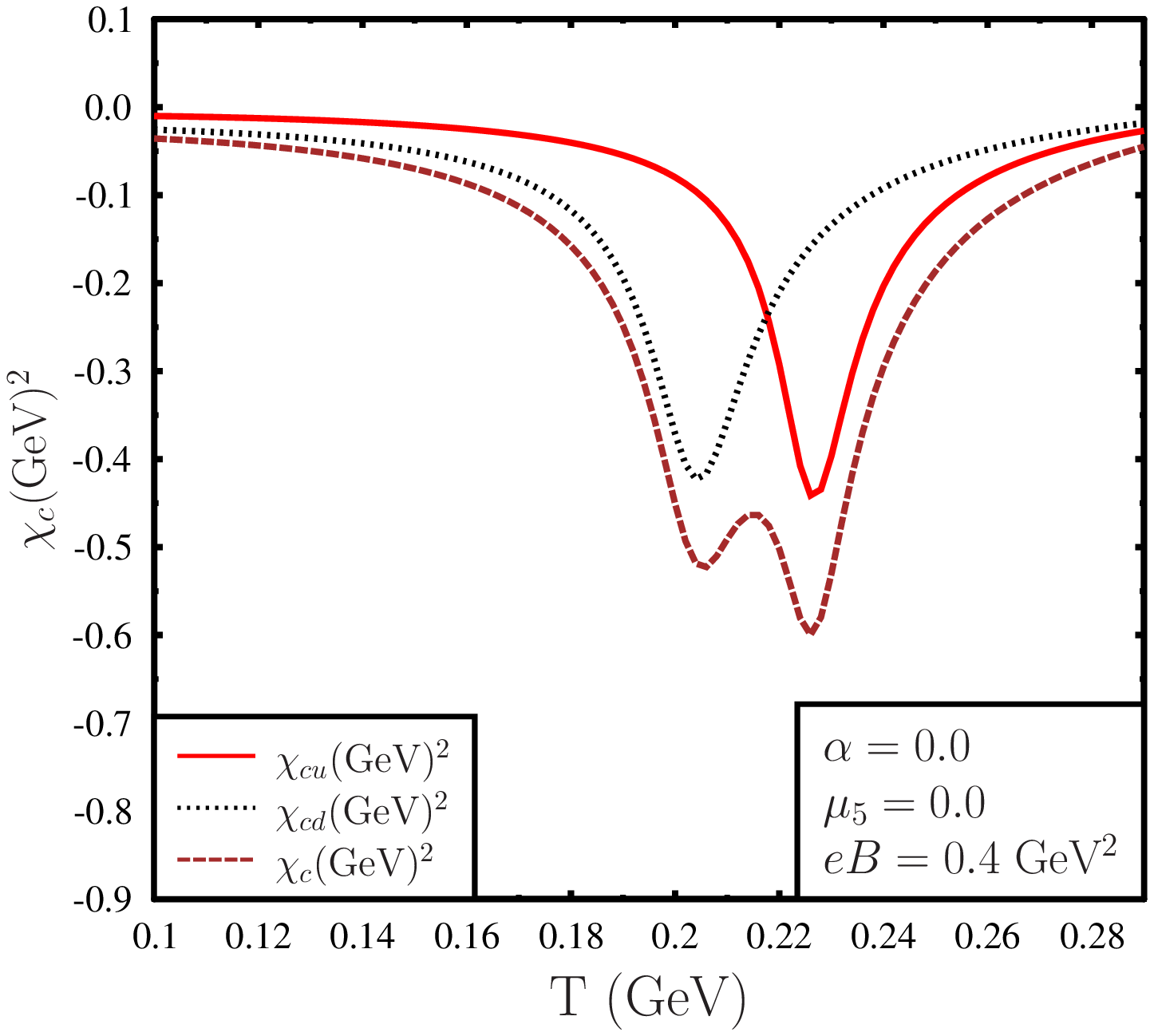}
    \end{minipage}%
    \begin{minipage}{0.5\textwidth}
        \centering
        \includegraphics[width=1.1\linewidth]{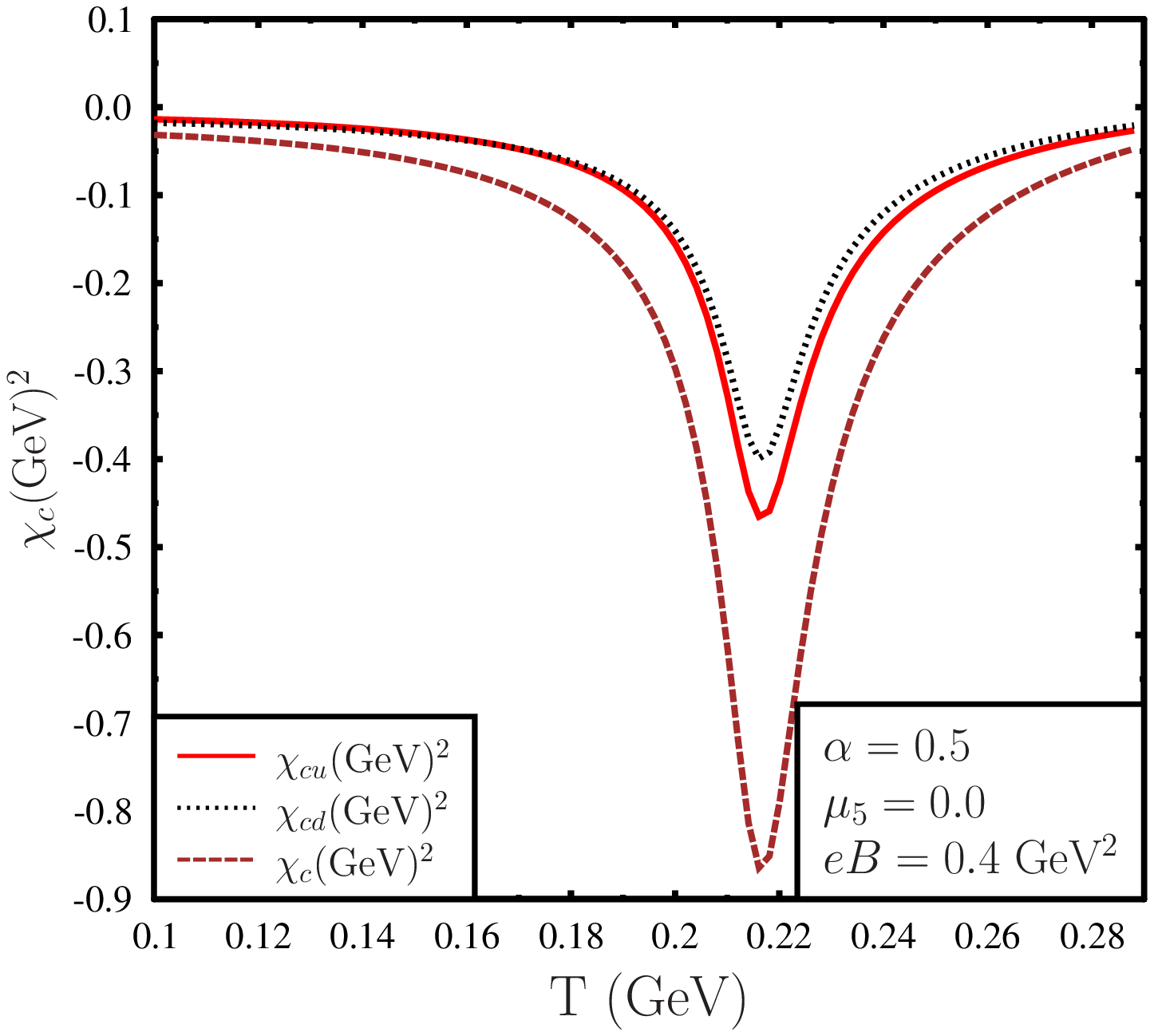}
    \end{minipage}
    \caption{Left plot: Variation of $\chi_{cu}$, $\chi_{cd}$ and $\chi_{c}$ with temperature at
    vanishing chiral chemical potential, for $eB=0.4$GeV$^2$ and $\alpha=0.0$.
    Right plot: Variation of $\chi_{cu}$, $\chi_{cd}$ and $\chi_{c}$ with temperature at
    vanishing chiral chemical potential, for $eB=0.4$GeV$^2$ and $\alpha=0.5$.
    From the left plot it is clear that chiral susceptibility shows two distinct peaks at large magnetic field. This is 
    due to the fact that at large magnetic field difference between $M_u$ and $M_d$ is large. On the other hand 
    right plot shows that, for $\alpha=0.5$, $\langle\bar{\psi}_u\psi_u\rangle\neq \langle\bar{\psi}_d\psi_d\rangle$, $\chi_{cu}$ and $\chi_{cd}$
      shows peak at same temperature. Hence for $\alpha=0.5$, at finite magnetic field chiral transition temperature 
    for $u$ and $d$ quarks are same.}
    \label{fig6}
\end{figure}

In Fig.\eqref{fig5} we show the variation of constituent quark masses $M_u$ and $M_d$ and the associated total chiral susceptibility,
with temperature for  vanishing chiral chemical potential and with different values of magnetic field for $\alpha=0.0$. For $\alpha = 0.0$ there is no flavour mixing. From the 
left plot it is clear that at finite magnetic field $M_u\neq M_d$. This is because in the presence of magnetic field 
$u$ and $d$ quark condensates are different and in the absence of flavour mixing for $\alpha=0.0$, $M_u$ is 
independent of $\langle\bar{\psi}_d\psi_d\rangle$. Similarly $M_d$ is 
independent of $\langle\bar{\psi}_u\psi_u\rangle$  for $\alpha=0.0$. From the right plot in Fig.\eqref{fig5} it is clear 
that chiral transition temperature increases with increasing magnetic field. However it is important to mention that 
unlike the case when $\alpha=0.5$, in this case susceptibility plot shows two distinct peaks at relatively large 
magnetic field values. In fact these two peaks are associated with $u$ and $d$ quarks, which has been shown in Fig.\eqref{fig6}.
In the left plot of Fig.\eqref{fig6} we show $\chi_{cu}$, $\chi_{cd}$ and $\chi_{c}$ for $eB=0.4$GeV$^2$ and $\alpha=0.0$. 
On the other hand In the right plot of Fig.\eqref{fig6} we show $\chi_{cu}$, $\chi_{cd}$ and $\chi_{c}$ for $eB=0.4$GeV$^2$
and $\alpha=0.5$. From the left plot in Fig.\eqref{fig6} it is clear that for $\alpha=0.0$, i.e. in the absence of flavour mixing,
at relatively large magnetic field chiral susceptibility $\chi_c$ shows two distinct peaks. These two peaks are associated 
with $u$ and $d$ quarks. At relatively large magnetic field with $\alpha=0.0$, chiral restoration of $d$ quark happens
at relatively low temperature with respect to the $u$ quarks. This is due to the fact that at non zero magnetic field 
$M_u>M_d$, as can be seen in Fig.\eqref{fig5}. On the other hand from the right plot in Fig.\eqref{fig6} we can see that,
although $\langle\bar{\psi}_u\psi_u\rangle\neq \langle\bar{\psi}_d\psi_d\rangle$,  $\chi_{cu}$ and $\chi_{cd}$
shows peak at same temperature. Hence for $\alpha=0.5$, at finite magnetic field chiral transition temperature
for $u$ and $d$ quarks are same.

\begin{figure}[!htb]
    \centering
    \begin{minipage}{.5\textwidth}
        \centering
        \includegraphics[width=1.1\linewidth]{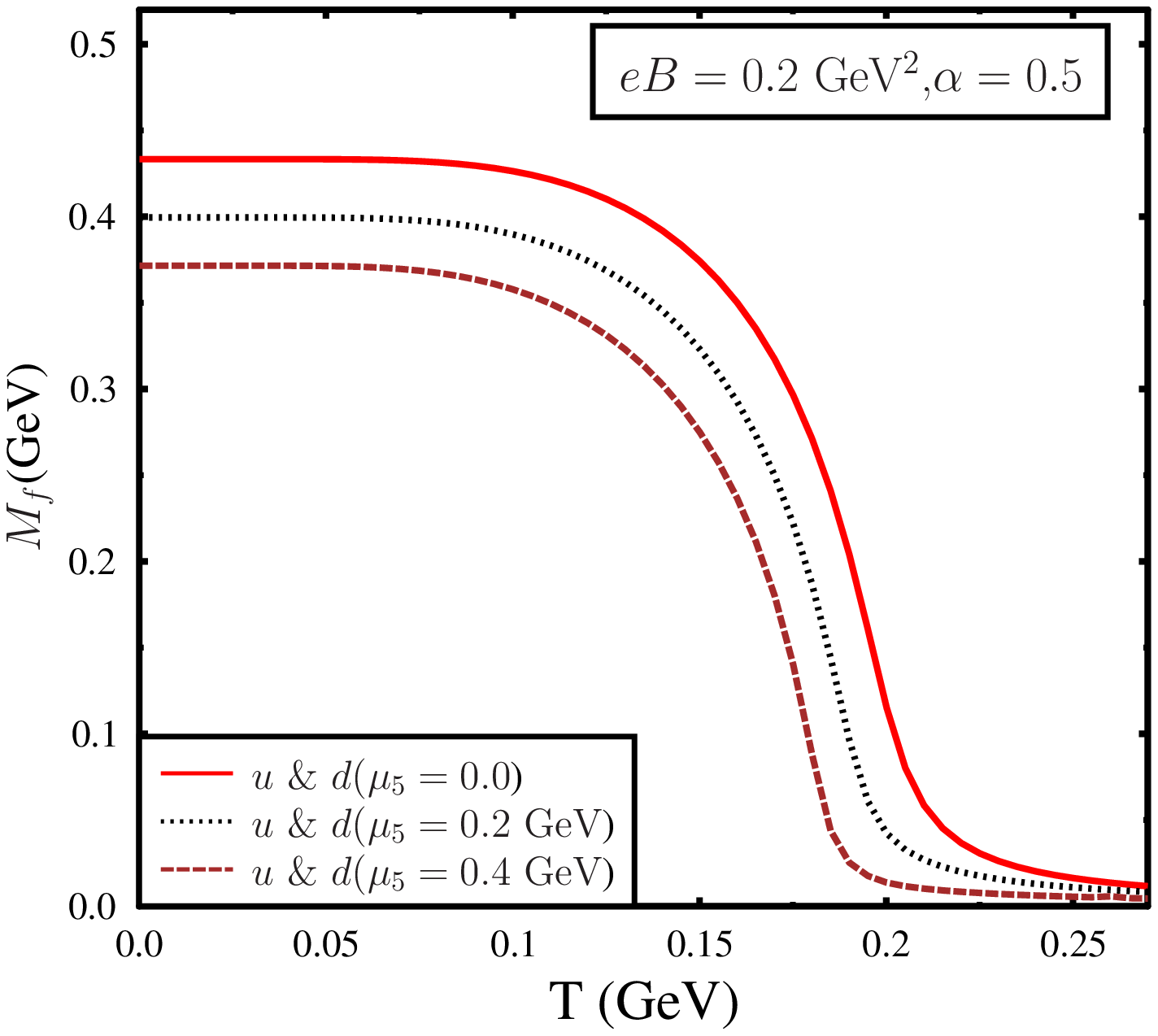}
    \end{minipage}%
    \begin{minipage}{0.5\textwidth}
        \centering
        \includegraphics[width=1.1\linewidth]{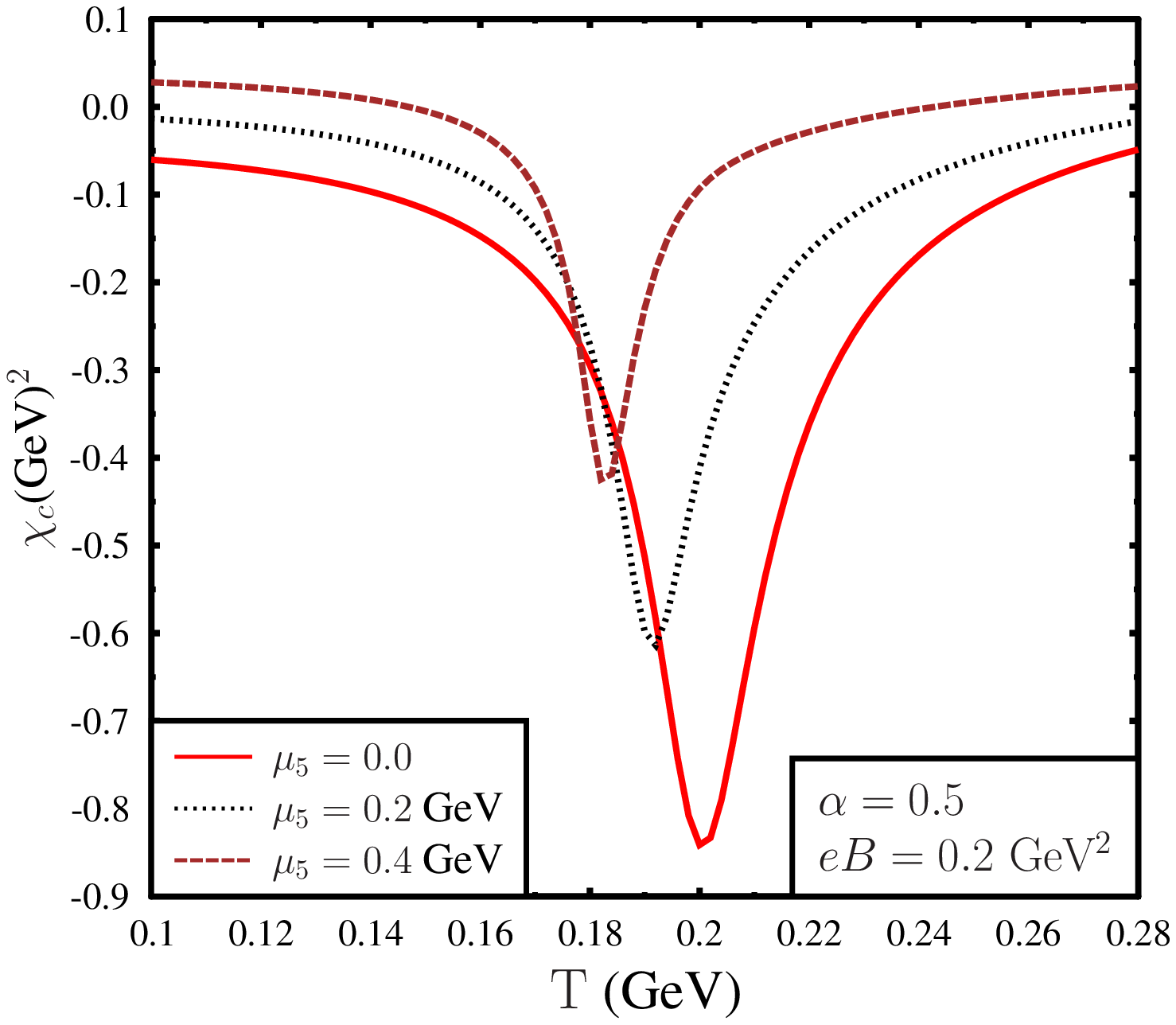}
    \end{minipage}
    \caption{Left plot: Variation of constituent quark mass $M_u=M_d$, with temperature
    for finite magnetic field and finite chiral chemical potential. Right Plot: Variation of chiral 
    susceptibility $\chi_c$ with temperature for finite magnetic field and finite chiral chemical potential. From this figure it is clear that with increasing
    chiral chemical potential quark mass as well as the chiral transition temperature decreases.}
    \label{fig7}
\end{figure}

 Finally in Fig.\eqref{fig7} we show the variation of quark constituent masses $M_u$ and $M_d$ and the associated
 susceptibilities with temperature for finite magnetic field and finite 
 chiral chemical potential for $\alpha=0.5$. Behaviour of quark constituent masses and the chiral susceptibilities with 
 temperature are similar for other values of $\alpha$. Left plot in Fig.\eqref{fig7} shows that with increasing value of 
 chiral chemical potential and for finite magnetic field constituent quark mass decreases. This decrease in mass with 
 increasing chiral chemical potential has also manifested in the right plot of Fig.\eqref{fig7}, which shows that with 
 increasing chiral chemical potential chiral transition temperature decreases.

 \section{conclusion}
 \label{conclu}
 In this investigation we have studied chiral phase transition and the associated chiral susceptibility of the medium 
 produced in ultra relativistic heavy ion collisions at vanishing quark chemical potential
 using Wigner function approach within the framework of two flavour NJL model. For a dynamical system 
 like the medium produced in heavy ion collision quantum effects can be relevant. Hence the quantum kinetic equation is a suitable 
 formalism to understand the evolution of these dynamical system. The central quantity of the quantum kinetic description is the 
 Wigner function. Wigner function is the quantum mechanical analogue of classical distribution function. Different components of Wigner function  
 satisfies quantum kinetic equation. However in this investigation we have restricted ourselves to the
 case of global equilibrium so that $T,\mu_5$ are constant and we do not consider evolution of Wigner function. In fact we 
 could have done the analysis by estimating the mean field thermodynamic potential and minimizing the same to get the 
 quark masses as well as the susceptibility.
 
 We have looked into the behaviour of quark masses and chiral susceptibility within a two flavour NJL model with flavour mixing determinant interaction.
 In the absence of magnetic field $u$ and $d$ quark masses are degenerate, due to the isospin symmetry. However 
 in the presence of magnetic field, due to different electric charge of $u$ and $d$ quark, 
 constituent mass of $u$ and $d$ quark can be different. Our results show that while flavour mixing instanton induced interaction 
 does not affect the quark masses in the absence of magnetic field, in the presence of magnetic field this
 interaction can affect quark masses. For maximal flavour mixing in NJL model for a non vanishing magnetic field $u$ and $d$
 quark masses are degenerate. For non maximal flavour mixing quark masses are non degenerate in the presence of magnetic field.
 Constituent mass of $u$ and $d$ quark is larger for non vanishing magnetic field compared to $B=0$ counterpart. With increasing 
 magnetic field constituent mass of $u$ and $d$ quark also increases. This apart, the chiral transition temperature is higher for 
 non vanishing magnetic field with respect to the vanishing magnetic field case. This is the manifestation of magnetic catalysis i.e. in the 
 presence of magnetic field the formation of chiral condensate is more probable, also the magnitude of the chiral condensate is higher for 
 larger magnetic field. It is important to note that in the presence of non maximal flavour mixing instanton interaction ,
 for vanishing magnetic field as well as for relatively small magnetic field the 
 the chiral transition temperatures of $u$ and $d$ quark are same. But for larger magnetic field transition temperature of $u$ quark and 
 $d$ quark are different. The difference between the transition temperature of $u$ and $d$ quark also increases with magnetic field. 
 We have also shown that, non vanishing chiral chemical potential ($\mu_5$) reduces quark mass in the absence as well as in the presence 
 of magnetic field. Unlike magnetic catalysis, with increasing chiral chemical potential ($\mu_5$), chiral transition temperature decreases.
 Also note that in presence of magnetic field, the chiral susceptibility shows a double peak structure due to isospin 
 breaking in presence of magnetic field.
 
 \section*{Acknowledgement}
 We thank Aman Abhishek for useful discussions on the medium seperation regularization scheme (MSS) used in this work 
 and also thank Jitesh R. Bhatt for useful discussions on Winger function formalism.
\appendix
\section{Derivation of scalar condensate in a background magnetic field and chiral chemical potential}
\label{appendix1}
Scalar condensate in the terms of the scalar DHW function can be written as,
\begin{align}
\langle \bar{\psi} \psi \rangle = \int d^4p \, F(p)
\label{appenA1}
\end{align}

Using the explicit form of scalar DHW function ($F(p)$) as given in Eq.\eqref{equ45}, scalar condensate in the presence
of magnetic field as given in Eq.\eqref{appenA1} can be expressed as,

\begin{align}
\langle \bar{\psi} \psi \rangle & = \int 2\pi p_T\,dp_0\,dp_T\,dp_z\,  M\bigg[ \sum_{n=0}^{\infty} V_n(p_0,p_z)\, \Lambda^{(n)}_{+}(p_T)\,\, 
+ \sum_{n=1}^{\infty} \frac{1}{\sqrt{p_z^2 + 2nqB}}\, A_n(p_0,p_z) p_z \,\Lambda^{(n)}_{-}(p_T) \bigg]\nonumber\\
& = \int 2\pi p_T\,dp_0\,dp_T\,dp_z\,  M\bigg[ V_0(p_0,p_z)\,
\Lambda^{(0)}_{+}(p_T)\, + \sum_{n=1}^{\infty} V_n(p_0,p_z)\, \Lambda^{(n)}_{+}(p_T)\, \nonumber\\
& ~~~~~~~~~~~~~~~~~~ ~~~~~~~~~~~~~~~~~~~~~~~~~~~~~~~~~~~~~~~~~~~~+ \sum_{n=1}^{\infty} 
\frac{1}{\sqrt{p_z^2 + 2nqB}}\, A_n(p_0,p_z) p_z \,\Lambda^{(n)}_{-}(p_T) \bigg]\nonumber\\
& = \mathbb{I}_1+\mathbb{I}_2+\mathbb{I}_3
\label{appenA2}
\end{align}

Now, the first term in Eq.\eqref{appenA2},
\begin{align}
 \mathbb{I}_1 = 2\,\pi\, \iint dp_0\,dp_z\,M V_0(p_0,p_z)\,\,\int dp_T\, p_T\, \Lambda_{+}^{(0)}(p_T).
 \label{appenA3}
\end{align}
Using the explicit form of $V_0(p_0,p_z)$ and $\Lambda_{+}^{(0)}(p_T)$, Eq.\eqref{appenA3} can be expressed as, 

\begin{align}
\mathbb{I}_1 &  = 2\,\pi\, \iint dp_0\,dp_z\, \frac{2}{(2\pi)^3}\, M \delta\bigg((p_0+\mu)^2-|E^{(0)}_{p_z}|^2\bigg)
\,\bigg[\theta(p_0+\mu)\,f_{FD}(p_0)\, \nonumber\\
& ~~~~~~~~~~~~~~~~~~~~~~~~~+  \theta(-p_0 -\mu)\,[f_{FD}(-p_0)-1]\bigg]\,\int dp_T\, p_T\, 2 \exp\left[{-p_T^2}/{qB}\right]
\nonumber\\
& = \frac{qB}{(2\pi)^2} \int dp_z\, \frac{M}{E^{(0)}_{p_z}} \bigg[ f_{FD}(E^{(0)}_{p_z}-\mu)\, 
+ f_{FD}(E^{(0)}_{p_z}+\mu)-1 \bigg]
\label{appenA4}
\end{align}

The second term in Eq.\eqref{appenA2},

\begin{align}
 \mathbb{I}_2 & = 2\pi \sum_{n=1}^{\infty}\iint dp_0~ dp_z~  M V_n(p_0,p_z)\int dp_T p_T \Lambda_{+}^{(n)}(p_T).
 \label{appenA5}
\end{align}

Using the explicit form of $\Lambda_{+}^{(n)}(p_T)$ one can calculate the following integral, 
\begin{align}
\int dp_T p_T \Lambda_{+}^{(n)}(p_T)& = (-1)^n\int_0^{\infty}dp_T~p_T \bigg[L_n(2p_T^2/qB)-L_{n-1}(2p_T^2/qB)\bigg]\exp(-p_T^2/qB)
= qB
\label{appenA6}
\end{align}
 To get the Eq.\eqref{appenA6} we use the following identity\cite{handbook}, 
 \begin{align}
  \int_0^{\infty}dx~\exp(-bx)L_n(x)=(b-1)^nb^{-n-1}.
  \label{appenA8}
 \end{align}

 Using Eq.\eqref{appenA6} and the explicit form of $V_n(p_0,p_z)$, $\mathbb{I}_2$ can be written as,
 
 \begin{align}
  \mathbb{I}_2 & =2\pi(qB)\iint~ dp_0~dp_z \frac{2}{(2\pi)^3}M\sum_s \delta\left((p_0+\mu)^2-|E^{(n)}_{p_z,s}|^2\right)
  \bigg[\theta(p_0+\mu)f_{FD}(p_0)+\theta(-p_0-\mu)(f_{FD}(-p_0)-1)\bigg]\nonumber\\
   & = -\frac{qB}{(2\pi)^2}\sum_{n=1}^{\infty}
  \sum_s\int dp_z \frac{M}{E_{pz,s}^{(n)}}
  \bigg[1-f_{FD}(E^{(n)}_{p_z,s}-\mu)-f_{FD}(E^{(n)}_{p_z,s}+\mu)\bigg]
\label{appenA9}
  \end{align}

Now let us consider the third term of Eq.\eqref{appenA2},

\begin{align}
\mathbb{I}_3 = 2\,\pi\,\iint dp_0\,dp_z\, M \sum_{n=1}^{\infty} \frac{1}{\sqrt{p_z^2 + 2nqB}}\, A_n(p_0,p_z)\, p_z \,\,\int dp_T\, p_T\, \Lambda_{-}^{(n)}(p_T).
\label{appenA10}
\end{align}
Using the explicit form of $\Lambda_{-}^{(n)}(p_T)$, it can be shown that,

\begin{align}
 \int~dp_T~p_T \Lambda_{-}^{(n)}(p_T) & = 0
 \label{appenA11}
\end{align}
Hence the third term of the Eq.\eqref{appenA2},
\begin{align}
 \mathbb{I}_3=0.
 \label{appenA13}
\end{align}

Hence using Eq.\eqref{appenA4}, Eq.\eqref{appenA9} and 
Eq.\eqref{appenA13}, the scalar condensate is,

\begin{align}
\langle \bar{\psi} \psi \rangle & = -\frac{qB}{(2\pi)^2} \int dp_z\, \frac{M}{E^{(0)}_{p_z}}
\bigg[1- f_{FD}(E^{(0)}_{p_z}-\mu)\, - f_{FD}(E^{(0)}_{p_z}+\mu) \bigg]\,\nonumber\\
& - \frac{qB}{(2\pi)^2}\sum_{n=1}^{\infty}\sum_s \int dp_z\,  
\frac{M}{E^{(n)}_{p_z,s}}\bigg[1-f_{FD}(E^{(n)}_{p_z,s}-\mu) - \,f_{FD}(E^{(n)}_{p_z,s}+\mu)\bigg].
\label{appenA14}
\end{align}

\section{Regularization of chiral condensate in a background magnetic field}
\label{appendix2}

The scalar condensate of a quark of flavour $f$, with $N_c$ color degrees of freedom at finite temperature $(T)$,
chemical potential ($\mu$) can be expressed as, 

\begin{align}
 \langle\bar{\psi}_f\psi_f\rangle^{\mu_5=0}_{B\neq0}  & =  -\frac{N_c|q_f|B}{(2\pi)^2}  \sum_{n=0}^{\infty} (2-\delta_{n,0})\int dp_z\,   
 \frac{M_{0_f}}{\epsilon^{(n)}_{p_z,f}}\Big[1-f_{FD}(\epsilon^{(n)}_{p_z,f}
 -\mu) - \,f_{FD}(\epsilon^{(n)}_{p_z,f}+\mu)\Big]\bigg]\nonumber\\
 & = \langle\bar{\psi}_f\psi_f\rangle^{\mu_5=0}_{vac,B\neq0}
 +\langle\bar{\psi}_f\psi_f\rangle^{\mu_5=0}_{med,B\neq0},
 \label{appenB1}
\end{align}
where $\langle\bar{\psi}_f\psi_f\rangle^{\mu_5=0}_{vac,B\neq0}$ is the $T=0, \mu=0$ part or the vacuum part of the scalar
condensate and
$\langle\bar{\psi}_f\psi_f\rangle^{\mu_5=0}_{med,B\neq0}$ is the finite temperature and finite
chemical potential part or the medium part of the scalar
condensate in the presence of magnetic field.
It is clear from the Eq.\eqref{appenB1} the vacuum term is divergent for large momenta and
however because of the distribution functions the medium part in Eq.\eqref{appenB1} is not.
Hence it is important to regulate the vacuum part in Eq.\eqref{appenB1}.

Let us consider the vacuum part $\langle\bar{\psi}_f\psi_f\rangle^{\mu_5=0}_{vac,B\neq0}$ which is 
given as,
\begin{align}
 \langle\overline{\psi}_f\psi_f\rangle^{\mu_5=0}_{vac,B\neq 0} & =-\frac{N_c}{2\pi}\sum_{n=0}^{\infty}(2-\delta_{n,0})|q_f|B
 \int_{-\infty}^{\infty}\frac{dp_z}{(2\pi)}\frac{M_{0_f}}{\epsilon_{p_z,f}^{(n)}}\nonumber\\
 & = -\frac{N_c}{2\pi}\sum_{n=0}^{\infty}2|q_f|B\int_{-\infty}^{\infty}\frac{dp_z}{(2\pi)}\frac{M_{0_f}}{\epsilon_{p_z,f}^{(n)}}
 +\frac{N_c}{2\pi}|q_f|B\int_{-\infty}^{\infty}\frac{dp_z}{(2\pi)}\frac{M_{0_f}}{\epsilon_{p_z,f}^{(0)}}\nonumber\\
 & = \mathcal{I}_1+\mathcal{I}_2
 \label{appenB2}
\end{align}

Both integrals $\mathcal{I}_1$ and $\mathcal{I}_2$ are divergent at large momentum. These integrals can be regularized using 
dimensional regularization scheme. In this regularization scheme integral $\mathcal{I}_1$ can be expressed as, 
 
\begin{align}
 \mathcal{I}_1 & = -\frac{N_c}{2\pi}\sum_{n=0}^{\infty}2|q_f|B\int_{-\infty}^{\infty}\frac{dp_z}{(2\pi)}\frac{M_{0_f}}{\epsilon_{p_z,f}^{(n)}}\nonumber\\
 &=-\frac{N_c}{2\pi}\sum_{n=0}^{\infty}2|q_f|B\frac{M_{0_f}
 \Gamma(\epsilon/2)}{(4\pi)^{(1-\epsilon)/2}\Gamma(1/2)(x_{0_f}+n)^{\epsilon/2}},
 \label{appenB6}
\end{align}
where the dimensionless variable $x_{0_f} \equiv M_{0_f}^2/2|q_f|B$.
Similarly the integral $\mathcal{I}_2$ can be expressed as,
\begin{align}
 \mathcal{I}_2 & =\frac{N_c}{2\pi}|q_f|B\int \frac{dp_z}{(2\pi)}\frac{M_{0_f}}{\sqrt{M_{0_f}^2+p_z^2}}\nonumber\\
 & = \frac{N_c M_{0_f} |q_f|B}{(2\pi)} \frac{\Gamma(\epsilon/2)}{(4\pi)^{(1-\epsilon)/2}\Gamma(1/2)x_{0_f}^{\epsilon/2}}
 \label{appenB7}
\end{align}

Using Eq.\eqref{appenB6} and Eq.\eqref{appenB7}, vacuum part of the scalar condensate in the presence of magnetic field as
given in Eq.\eqref{appenB2}, can be recasted as,

\begin{align}
 \mathcal{I}_1+\mathcal{I}_2 & =-\frac{N_c}{2\pi}2|q_f|BM_{0_f}\frac{\Gamma(\epsilon/2)}{(4\pi)^{(1-\epsilon)/2}
 \Gamma(1/2)}\bigg[\sum_{n=0}^{\infty}\frac{1}{(x_{0_f}+n)^{\epsilon/2}}-\frac{1}{2x_{0_f}^{\epsilon/2}}\bigg]\nonumber\\
 & = -\frac{N_c}{2\pi}2|q_f|BM_{0_f}\frac{\Gamma(\epsilon/2)}{(4\pi)^{1/2}\Gamma(1/2)}\bigg[\zeta(\epsilon/2,x_{0_f})
 -\frac{1}{2x_{0_f}^{\epsilon/2}}\bigg].
 \label{appenB8}
 \end{align}
 Expanding the right hand side of Eq.\eqref{appenB8} around $\epsilon\rightarrow 0$ and keeping only the leading order terms, we get,
 \begin{align}
 \mathcal{I}_1+\mathcal{I}_2 
& = -\frac{N_c}{2\pi^2}|q_f|BM_{0_f} \bigg[-\frac{2x_{0_f}}{\epsilon}+\gamma_E x_{0_f}+\frac{1}{2}\ln x_{0_f}
+\ln \Gamma(x_{0_f})-\frac{1}{2}\ln(2\pi)\bigg]
 \label{appenB9}
 \end{align}
In Eq.\eqref{appenB8}, we have used the representation of Zeta function, which is given as \cite{zetafunction},
\begin{align}
 \zeta(a,x)=\sum_{n=0}^{\infty}\frac{1}{(x+n)^a},
 \label{appenB10}
\end{align}
also, we have used the following identities to get 
Eq.\eqref{appenB9},

\begin{align}
 \zeta(0,x)=(\frac{1}{2}-x),~~\text{and},~~
 \zeta^{\prime}(0,x)=\ln \Gamma(x)-\frac{1}{2}\ln(2\pi),~~\text{where}~~\zeta^{\prime}(0,x)=\frac{d\zeta(a,x)}{da}|_{a=0}
 \label{appenB11}
\end{align}
It is clear from Eq.\eqref{appenB9}, that the vacuum
part  has $1/\epsilon$ divergent part. To remove this $1/\epsilon$ divergence 
we use the following integral,

 \begin{align}
 \mathcal{I}_3 =-2N_c\int\frac{d^3p}{(2\pi)^3}\frac{M_{0_f}}{\sqrt{p^2+M_{0_f}^2}}
  \label{appenB12}
  \end{align}
  Using dimensional regularization method the integral in Eq.\eqref{appenB12} can be recasted as,
  \begin{align}
 \mathcal{I}_3
  & = \frac{-2N_c M_{0_f}}{(4\pi)^{3/2}\Gamma(1/2)}\frac{\Gamma(-1+\epsilon/2)}{(2x_{0_f}|q_f|B)^{-1+\epsilon/2}}
  \label{appenB13}
  \end{align}
  Expand the right hand side of Eq.\eqref{appenB13} 
  around $\epsilon\rightarrow 0$ and keeping only the leading order terms we get,
  \begin{align}
  \mathcal{I}_3 = \frac{-N_c M_{0_f} |q_f|B}{2\pi^2}\bigg[-\frac{2x_{0_f}}{\epsilon}-x_{0_f}+x_{0_f}\gamma_E+x_{0_f}\ln x_{0_f}\bigg]
  \label{appenB14}
 \end{align}

 Using Eq.\eqref{appenB9} and Eq.\eqref{appenB14} we get, 
 \begin{align}
  \mathcal{I}_1+\mathcal{I}_2-\mathcal{I}_3 
  & = -\frac{N_c M_{0_f} |q_f|B}{2\pi^2}\bigg[x_{0_f}(1-\ln x_{0_f})+\ln \Gamma(x_{0_f})+\frac{1}{2}\ln 
  \bigg(\frac{x_{0_f}}{2\pi}\bigg)\bigg]
  \label{appenB15}
 \end{align}

 Using Eq.\eqref{appenB2} and Eq.\eqref{appenB15}, we have the regularized vacuum part of the scalar condensate in the 
 presence of magnetic field and is given as,
 \begin{align}
\langle\overline{\psi}_f\psi_f\rangle^{\mu_5=0}_{vac,B\neq 0} & = \mathcal{I}_1+\mathcal{I}_2-\mathcal{I}_3+\mathcal{I}_3 \nonumber\\
& = -\frac{N_c M_{0_f} |q_f|B}{2\pi^2}\bigg[x_{0_f}(1-\ln x_{0_f})+\ln \Gamma(x_{0_f})+\frac{1}{2}\ln 
\bigg(\frac{x_{0_f}}{2\pi}\bigg)\bigg]
-2N_c\int_{|\vec{p}|\leq \Lambda}\frac{d^3p}{(2\pi)^3}\frac{M_{0_f}}{\sqrt{p^2+M_{0_f}^2}}.
\label{appenB16}
 \end{align}
 
 Again,  
 \begin{align}
  \mathcal{I}_1 &  = \mathcal{I}_1-\mathcal{I}_3+\mathcal{I}_3\nonumber\\
  & = -\frac{N_c|q_f|B M_{0_f}}{2\pi^2}\Bigg[\frac{1}{\epsilon}-\frac{\gamma_E}{2}+x_{0_f}(1-\ln x_{0_f})+\ln \Gamma(x_{0_f})
 -\frac{1}{2}\ln(2\pi)\Bigg]-2N_c \int \frac{d^3p}{(2\pi)^3}\frac{M_{0_f}}{\sqrt{p^2+M_{0_f}^2}}.
 \label{appenB17}
 \end{align}
Hence,
 \begin{align}
 \mathcal{I} & \equiv -\frac{N_c|q_f|B}{(2\pi)^2}\sum_{s=\pm 1}\sum_{n=0}^{\infty}\int dp_z \frac{1}{\sqrt{p_z^2+M_{0_f}^2+2n|q_f|B}}
 \nonumber\\ 
 & = 
 -\frac{N_c}{2\pi^2}|q_f|B\Bigg[\frac{1}{\epsilon}-\frac{\gamma_E}{2}+x_{0_f}(1-\ln x_{0_f})+\ln \Gamma(x_{0_f})
 -\frac{1}{2}\ln(2\pi)\Bigg] -2N_c \int \frac{d^3p}{(2\pi)^3}\frac{1}{\sqrt{p^2+M_{0_f}^2}}
 \label{appenB18}
\end{align}
Using Eq.\eqref{appenB18} we get, 
\begin{align}
 & \frac{N_c|q_f|B}{(2\pi)^2}\sum_{s=\pm 1}\sum_{n=0}^{\infty}\int dp_z \frac{1}{(p_z^2+M_{0_f}^2+2n|q_f|B)^{3/2}}\equiv 
  \frac{1}{M_{0_f}}\frac{\partial \mathcal{I}}{\partial M_{0_f}} \nonumber\\
  & = -\frac{N_c}{2\pi^2}\Bigg[-\ln x_{0_f}+\frac{\Gamma^{\prime}(x_{0_f})}{\Gamma(x_{0_f})}\Bigg] +2N_c\int_{|\vec{p}|\leq\Lambda}
  \frac{d^3p}{(2\pi)^3}\frac{1}{(p^2+M_{0_f}^2)^{3/2}}
   \label{appenB19}
\end{align}

 \section{Regularization of chiral condensate in a background magnetic field and chiral chemical potential}
\label{appendix3}

The scalar condensate of a quark of flavour $f$ with $N_c$ color degrees of freedom at finite temperature $(T)$,
quark chemical potential ($\mu$), chiral chemical potential ($\mu_5$), electric charge ($q_f$) and magnetic field ($B$) can 
be expressed as, 

\begin{align}
\langle \bar{\psi}_f \psi_f \rangle^{\mu_5\neq0}_{B\neq0} & = -\frac{N_c|q_f|B}{(2\pi)^2} \bigg[\int dp_z\, \frac{M_f}{E^{(0)}_{p_z,f}} 
\Big[ 1-f_{FD}(E^{(0)}_{p_z,f}-\mu)\, - 
f_{FD}(E^{(0)}_{p_z,f}+\mu)\Big]\,\nonumber\\
 & +  \sum_{n=1}^{\infty}\sum_s \int dp_z\,   \frac{M_f}{E^{(n)}_{p_z,s,f}}\Big[1-f_{FD}(E^{(n)}_{p_z,s,f}-\mu) 
 - \,f_{FD}(E^{(n)}_{p_z,s,f}+\mu)\Big]\bigg]\nonumber\\
 & = \langle\bar{\psi}_f\psi_f\rangle^{\mu_5\neq0}_{vac,B\neq0}
 +\langle\bar{\psi}_f\psi_f\rangle^{\mu_5\neq0}_{med,B\neq0},
 \label{appenC1}
 \end{align}
where $\langle\bar{\psi}_f\psi_f\rangle^{\mu_5\neq0}_{vac,B\neq0}$ is the $T=0, \mu=0$ part or the vacuum part of the scalar
condensate and 
$\langle\bar{\psi}_f\psi_f\rangle^{\mu_5\neq0}_{med,B\neq0}$ is the finite temperature and finite
chemical potential part or the medium part of the scalar
condensate in the presence of magnetic field and chiral chemical potential ($\mu_5$). It is clear from the Eq.\eqref{appenC1} 
that the vacuum term is divergent at large momenta and
however because of the distribution functions the medium part in Eq.\eqref{appenC1} is not.
Hence the vacuum term has to be regularized.

The vacuum term in the presence of magnetic field and chiral chemical potential can be expressed as, 

\begin{align}
 \langle\bar{\psi}_f\psi_f\rangle^{\mu_5\neq0}_{vac,B\neq0}& = -N_c\frac{|q_f|B}{(2\pi)^2}\int dp_z \frac{M_f}{\sqrt{M_f^2+(p_z-\mu_5)^2}}
  -N_c\frac{|q_f|B}{(2\pi)^2}\sum_{n=1}^{\infty}\sum_{s=\pm1}\int dp_z \frac{M_f}{\sqrt{M_f^2+(\sqrt{p_z^2+2n|q_f|B}-s\mu_5)^2}}\nonumber\\
 & = - N_c\frac{|q_f|B}{(2\pi)^2}\sum_{n=0}^{\infty}\sum_{s=\pm1}\int dp_z \frac{M_f}{\sqrt{M_f^2+(\sqrt{p_z^2+2n|q_f|B}-s\mu_5)^2}}
 +N_c\frac{|q_f|B}{(2\pi)^2}\int dp_z \frac{M_f}{\sqrt{M_f^2+(p_z-\mu_5)^2}}\nonumber\\
 & = - N_c\frac{|q_f|B}{(2\pi)^2}\sum_{n=0}^{\infty}\sum_{s=\pm1}\int dp_z \frac{1}{\pi}\int_{-\infty}^{\infty}
 dp_4\frac{M_f}{p_4^2+\left(M_f^2+(\sqrt{p_z^2+2n|q_f|B}-s\mu_5)^2\right)}\nonumber\\
 & ~~~~~~~~~~~~~~~~~~~~ +N_c\frac{|q_f|B}{(2\pi)^2}\int dp_z \frac{1}{\pi} \int_{-\infty}^{\infty} dp_4
 \frac{M_f}{p_4^2+M_f^2+(p_z-\mu_5)^2}\nonumber\\
 & = I_1+I_2
 \label{appenC2}
\end{align}

Using the regularization method discussed in Ref.\cite{chiralNJL7} we can write the integrand of the integral $I_1$ as given
in the Eq.\eqref{appenC2} as following  

\begin{align}
 & \frac{1}  {p_4^2+M_f^2+(\sqrt{p_z^2+2n|q_f|B}-s\mu_5)^2}\nonumber\\
 &= \frac{1}{p_4^2+p_z^2+M_{0_f}^2+2n|q_f|B}
 -\frac{1}{p_4^2+p_z^2+M_{0_f}^2+2n|q_f|B}+\frac{1}  {p_4^2+M_f^2+(\sqrt{p_z^2+2n|q_f|B}-s\mu_5)^2}\nonumber\\
 & = \frac{1}{p_4^2+p_z^2+M_{0_f}^2+2n|q_f|B} +\frac{M_{0_f}^2-M_f^2-\mu_5^2+2s\mu_5\sqrt{p_z^2+2n|q_f|B}}
 {\bigg(p_4^2+p_z^2+M_{0_f}^2+2n|q_f|B\bigg)
 \bigg(p_4^2+M_f^2+(\sqrt{p_z^2+2n|q_f|B}-s\mu_5)^2\bigg)}
 \label{appenC3}
 \end{align}
Using Eq.\eqref{appenC3} twice we can write the integrand of the integral $I_1$ in the 
following way, 

\begin{align}
 & \frac{1}{p_4^2+M_f^2+(\sqrt{p_z^2+2n|q_f|B}-s\mu_5)^2} = \frac{1}{p_4^2+p_z^2+M_{0_f}^2+2n|q_f|B} 
 + \frac{A+2s\mu_5\sqrt{p_z^2+2n|q_f|B}}{\bigg(p_4^2+p_z^2+M_{0_f}^2+2n|q_f|B\bigg)^2}\nonumber\\
 & ~~~~~~~~~~~~ + \frac{(A+2s\mu_5\sqrt{p_z^2+2n|q_f|B})^2}{\bigg(p_4^2+p_z^2+M_{0_f}^2+2n|q_f|B\bigg)^3}
 +\frac{(A+2s\mu_5\sqrt{p_z^2+2n|q_f|B})^3}{\bigg(p_4^2+p_z^2+M_{0_f}^2+2n|q_f|B\bigg)^3
 \bigg(p_4^2+M_f^2+(\sqrt{p_z^2+2n|q_f|B}-s\mu_5)^2\bigg)}, 
 \label{appenC4}
\end{align}
where $A=M_{0_f}^2-M_f^2-\mu_5^2$. Performing $p_4$ integration in each term of Eq.\eqref{appenC4} we get,

\begin{align}
 \frac{1}{\pi}\sum_s\int dp_4 \frac{1}{p_4^2+p_z^2+M_{0_f}^2+2n|q_f|B} = \sum_s\frac{1}{\sqrt{p_z^2+M_{0_f}^2+2n|q_f|B}} 
 \label{appenC5}
\end{align}
\begin{align}
  \frac{1}{\pi}\sum_s\int dp_4 \frac{A+2s\mu_5\sqrt{p_z^2+2n|q_f|B}}{\bigg(p_4^2+p_z^2+M_{0_f}^2+2n|q_f|B\bigg)^2}
  & = \sum_s \frac{1}{2} \frac{A}{\bigg(p_z^2+M_{0_f}^2+2n|q_f|B\bigg)^{3/2}}
  \label{appenC6}
\end{align}
\begin{align}
 & \frac{1}{\pi}\sum_s\int dp_4 \frac{(A+2s\mu_5\sqrt{p_z^2+2n|q_f|B})^2}{\bigg(p_4^2+p_z^2+M_{0_f}^2+2n|q_f|B\bigg)^3} 
  = \sum_s\Bigg[\frac{3}{8}\frac{A^2}{(p_z^2+M_{0_f}^2+2n|q_f|B)^{5/2}}
  -\frac{3}{2}\frac{\mu_5^2M_{0_f}^2}{(p_z^2+M_{0_f}^2+2n|q_f|B)^{5/2}}\nonumber\\
  & ~~~~~~~~~~~~~~~~~~~~~~~~~~~~~~~~~~~~~~~~~~~~~~~~~~~~~~~~~~~~~~~+\frac{3}{2}\frac{\mu_5^2}{(p_z^2+M_{0_f}^2+2n|q_f|B)^{3/2}}\Bigg]
  \label{appenC7}
\end{align}
\begin{align}
 & \frac{1}{\pi}\sum_s\int dp_4 \frac{(A+2s\mu_5\sqrt{p_z^2+2n|q_f|B})^3}{\bigg(p_4^2+p_z^2+M_{0_f}^2+2n|q_f|B\bigg)^3
 \bigg(p_4^2+M_f^2+(\sqrt{p_z^2+2n|q_f|B}-s\mu_5)^2\bigg)}\nonumber\\
 & = \frac{1}{\pi}\sum_s\int dp_4 \int_0^1 dx  \frac{3(1-x)^2(A+2s\mu_5\sqrt{p_z^2+2n|q_f|B})^3}
 {\Bigg[x\bigg(p_4^2+M_f^2+(\sqrt{p_z^2+2n|q_f|B}-s\mu_5)^2\bigg)+(1-x)\bigg(p_4^2+p_z^2+M_{0_f}^2+2n|q_f|B\bigg)\Bigg]^4}\nonumber\\
  &  = \sum_s \frac{15}{16}\int_0^1 dx \frac{(1-x)^2(A+2s\mu_5\sqrt{p_z^2+2n|q_f|B})^3}
 {\Bigg[p_z^2+M_{0_f}^2+2n|q_f|B-x(A+2s\mu_5\sqrt{p_z^2+2n|q_f|B})\Bigg]^{7/2}}
 \label{appenC8}
\end{align}

Using Eq.\eqref{appenC5},Eq.\eqref{appenC6},Eq.\eqref{appenC7} and Eq.\eqref{appenC8}, integral $I_1$ in Eq.\eqref{appenC2} can be 
expressed as, 

\begin{align}
 I_1 & =  - N_c\frac{|q_f|B}{(2\pi)^2}\sum_{n=0}^{\infty}\sum_{s=\pm1}\int dp_z \frac{M_f}{\sqrt{M_f^2+
 (\sqrt{p_z^2+2n|q_f|B}-s\mu_5)^2}}\nonumber\\
& = I_{1_{\text{quad}}} -\frac{M_f(M_{0_f}^2-M_f^2+2\mu_5^2)}{2}I_{1_{\text{log}}}+I_{1_{\text{finite1}}}+I_{1_{\text{finite2}}},
 \label{appenC9}
  \end{align}
  where
  \begin{align}
   I_{1_{\text{quad}}} 
   = - N_c\frac{|q_f|B}{(2\pi)^2}\sum_{n=0}^{\infty}\sum_{s=\pm1}\int dp_z \frac{M_f}{\sqrt{p_z^2+M_{0_f}^2+2n|q_f|B}},
  \end{align}
  \begin{align}
   I_{1_{\text{log}}}= N_c\frac{|q_f|B}{(2\pi)^2}\sum_{n=0}^{\infty}\sum_{s=\pm1}\int dp_z
\frac{1}{(p_z^2+M_{0_f}^2+2n|q_f|B)^{3/2}},
  \end{align}
  \begin{align}
   I_{1_{\text{finite1}}}=- N_c\frac{|q_f|B}{(2\pi)^2}\sum_{n=0}^{\infty}\sum_{s=\pm1}\int dp_z  \left(\frac{3}{8}\right) 
\Bigg[\frac{ M_f A^2-4M_fM_{0_f}^2\mu_5^2}{(p_z^2+M_{0_f}^2+2n|q_f|B)^{5/2}}\Bigg],
  \end{align}
\begin{align}
 I_{1_{\text{finite2}}}= - N_c\frac{|q_f|B}{(2\pi)^2} \left(\frac{15}{16}\right) \sum_{n=0}^{\infty}\sum_{s=\pm1}\int dp_z 
 \int_0^1 dx \frac{(1-x)^2M_f(A+2s\mu_5\sqrt{p_z^2+2n|q_f|B})^3}
 {\Bigg[p_z^2+M_{0_f}^2+2n|q_f|B-x(A+2s\mu_5\sqrt{p_z^2+2n|q_f|B})\Bigg]^{7/2}}.
\end{align}

In a similar way the integral $I_2$ in Eq.\eqref{appenC2} can also be written as, 
\begin{align}
I_2 & = N_c\frac{|q_f|B}{(2\pi)^2}\int dp_z \frac{M_f}{\sqrt{M_f^2+(p_z-\mu_5)^2}}\nonumber\\
& = N_c\frac{|q_f|B}{(2\pi)^2}\int dp_z \frac{M_f}{\sqrt{M_f^2+(p_z-\mu_5)^2}}
-N_c\frac{|q_f|B}{(2\pi)^2}\int dp_z \frac{M_f}{\sqrt{p_z^2+M_{0_f}^2}}+N_c\frac{|q_f|B}{(2\pi)^2}\int dp_z 
\frac{M_f}{\sqrt{p_z^2+M_{0_f}^2}}\nonumber\\
& = \left(\frac{1}{2}\right)
 N_c\frac{|q_f|B}{(2\pi)^2}\int dp_z \int_0^1 dx \frac{M_f\bigg(A+2p_z\mu_5\bigg)}{\Bigg[p_z^2+M_{0_f}^2
 -x\bigg(A+2p_z\mu_5\bigg)\Bigg]^{3/2}}
+N_c\frac{|q_f|B}{(2\pi)^2}\int dp_z \frac{M_f}{\sqrt{p_z^2+M_{0_f}^2}}\nonumber\\
& = I_{2_{\text{finite}}}+I_{2_{\text{log}}}
\label{appenC10}
 \end{align}
 
 Using Eq.\eqref{appenC9} and Eq.\eqref{appenC10}, Eq.\eqref{appenC2} can be recasted as, 
 
 \begin{align}
  \langle\bar{\psi}_f\psi_f\rangle^{\mu_5\neq0}_{vac,B\neq0} = -\frac{M_f(M_{0_f}^2-M_f^2+2\mu_5^2)}{2}I_{1_{\text{log}}}
  +I_{1_{\text{finite1}}}+I_{1_{\text{finite2}}}+I_{2_{\text{finite}}}+I_{1_{\text{quad}}}+I_{2_{\text{log}}},
 \end{align}
where
\begin{align}
 I_{1_{\text{log}}} = -\frac{N_c}{2\pi^2}\Bigg[-\ln x_{0_f}+\frac{\Gamma^{\prime}(x_{0_f})}{\Gamma(x_{0_f})}\Bigg]
 +2N_c\int_{|\vec{p}|\leq\Lambda}\frac{d^3p}{(2\pi)^3}\frac{1}{(p^2+M_{0_f}^2)^{3/2}},
 \label{appenC16}
\end{align}
and 
\begin{align}
 I_{1_{\text{quad}}}+I_{2_{\text{log}}}& = -\frac{N_c M_{f} |q_f|B}{2\pi^2}\bigg[x_{0_f}(1-\ln x_{0_f})+\ln \Gamma(x_{0_f})+\frac{1}{2}\ln 
\bigg(\frac{x_{0_f}}{2\pi}\bigg)\bigg]-2N_c\int_{|\vec{p}|\leq\Lambda}\frac{d^3p}{(2\pi)^3}\frac{M_{f}}{\sqrt{p^2+M_{0_f}^2}}.
\label{appenC17}
\end{align}
In Eq.\eqref{appenC16} and \eqref{appenC17} we have used \eqref{appenB19} and \eqref{appenB16} respectively.

\section{Chiral susceptibility and its regularization in the presence of a background magnetic field and 
chiral chemical potential}
\label{appendix4}

Using Eq.\eqref{appenC2}, we get,
\begin{align}
 \frac{\partial\langle\bar{\psi}_f\psi_f\rangle^{\mu_5\neq0}_{vac,B\neq0}}{\partial M_f} & = -\frac{N_c|q_f|B}{(2\pi)^2}\sum_{n=0}^{\infty}\sum_{s=\pm1}
 \int dp_z \frac{1}{\sqrt{M_f^2+(\sqrt{p_z^2+2n|q_f|B}-s\mu_5)^2}}
 + \frac{N_c|q_f|B}{(2\pi)^2} \int dp_z \frac{1}{\sqrt{M_f^2+(p_z-\mu_5)^2}}\nonumber\\
& +\frac{N_c|q_f|B}{(2\pi)^2}\sum_{n=0}^{\infty}\sum_{s=\pm1}
 \int dp_z \frac{M_f^2}{\left(M_f^2+(\sqrt{p_z^2+2n|q_f|B}-s\mu_5)^2\right)^{3/2}}
 - \frac{N_c|q_f|B}{(2\pi)^2} \int dp_z \frac{M_f^2}{\left(M_f^2+(p_z-\mu_5)^2\right)^{3/2}}\nonumber\\
 & = \mathbf{I}_1 + \mathbf{I}_2+\mathbf{I}_3+\mathbf{I}_4
 \label{appenD1}
\end{align}

Using Eq.\eqref{appenC9}, we can write, 
\begin{align}
\mathbf{I}_1= & - N_c\frac{|q_f|B}{(2\pi)^2}\sum_{n=0}^{\infty}\sum_{s=\pm1}\int dp_z \frac{1}{\sqrt{M_f^2+
 (\sqrt{p_z^2+2n|q_f|B}-s\mu_5)^2}}\nonumber\\
 & = \mathbf{I}_{1,\text{quad}}-\frac{(M_{0_f}^2-M_f^2+2\mu_5^2)}{2}\mathbf{I}_{1,\text{log}}+\mathbf{I}_{1,\text{finite1}}
+\mathbf{I}_{1,\text{finite2}},
\label{appenD2}
  \end{align}
where
\begin{align}
 \mathbf{I}_{1,\text{quad}}
 = - N_c\frac{|q_f|B}{(2\pi)^2}\sum_{n=0}^{\infty}\sum_{s=\pm1}\int dp_z \frac{1}{\sqrt{p_z^2+M_{0_f}^2+2n|q_f|B}},
 \label{appenD3}
\end{align}
\begin{align}
 \mathbf{I}_{1,\text{log}}= N_c\frac{|q_f|B}{(2\pi)^2}\sum_{n=0}^{\infty}\sum_{s=\pm1}\int dp_z 
\frac{1}{(p_z^2+M_{0_f}^2+2n|q_f|B)^{3/2}},
\label{appenD4}
\end{align}
\begin{align}
 \mathbf{I}_{1,\text{finite1}} = 
 - N_c\frac{|q_f|B}{(2\pi)^2}\sum_{n=0}^{\infty}\sum_{s=\pm1}\int dp_z \bigg(\frac{3}{8}\bigg)
 \Bigg[\frac{ A^2-4M_{0_f}^2\mu_5^2}{(p_z^2+M_{0_f}^2
+2n|q_f|B)^{5/2}}\Bigg],
\label{appenD5}
\end{align}
\begin{align}
 \mathbf{I}_{1,\text{finite2}}= - N_c\frac{|q_f|B}{(2\pi)^2} \left(\frac{15}{16}\right) \sum_{n=0}^{\infty}\sum_{s=\pm1}\int dp_z 
 \int_0^1 dx\frac{(1-x)^2(A+2s\mu_5\sqrt{p_z^2+2n|q_f|B})^3}
 {\Bigg[p_z^2+M_{0_f}^2+2n|q_f|B-x(A+2s\mu_5\sqrt{p_z^2+2n|q_f|B})\Bigg]^{7/2}}.
 \label{appenD6}
\end{align}

The integral $\mathbf{I}_2$ in Eq.\eqref{appenD1} can be expressed as, 
\begin{align}
\mathbf{I}_2 & = N_c\frac{|q_f|B}{(2\pi)^2}\int dp_z \frac{1}{\sqrt{M_f^2+(p_z-\mu_5)^2}}
= \mathbf{I}_{2,\text{finite}}+\mathbf{I}_{2,\text{log}},
\label{appenD7}
 \end{align}
where divergence free $\mathbf{I}_{2,\text{finite}}$ is,
\begin{align}
\mathbf{I}_{2,\text{finite}}
 = \left(\frac{1}{2}\right)
 N_c\frac{|q_f|B}{(2\pi)^2}\int dp_z \int_0^1 dx \frac{\bigg(A+2p_z\mu_5\bigg)}{\Bigg[p_z^2+M_{0_f}^2
 -x\bigg(A+2p_z\mu_5\bigg)\Bigg]^{3/2}},
 \label{appenD8}
 \end{align}
and the divergence term $\mathbf{I}_{2,\text{log}}$ is, 
\begin{align}
 \mathbf{I}_{2,\text{log}}  = \frac{N_c|q_f|B}{(2\pi)^2}\int dp_z \frac{1}{\sqrt{p_z^2+M_{0_f}^2}}.
\end{align}

Similarly, the integral $\mathbf{I}_3$ can be separated into a divergent and a convergent terms as, 
\begin{align}
 \mathbf{I}_3 & = \frac{N_c|q_f|B}{(2\pi)^2}\sum_{n=0}^{\infty}\sum_{s=\pm1}
 \int dp_z \frac{M_f^2}{\left(M_f^2+(\sqrt{p_z^2+2n|q_f|B}-s\mu_5)^2\right)^{3/2}}
 = \mathbf{I}_{3,\text{finite}}+\mathbf{I}_{3,\text{log}},
 \label{appenD9}
 \end{align}
 where
 \begin{align}
  \mathbf{I}_{3,\text{finite}} =
  \frac{N_c|q_f|B}{(2\pi)^2}\sum_{n=0}^{\infty}\sum_{s=\pm1}
 \int dp_z M_f^2\Bigg[\frac{1}{\left(M_f^2+(\sqrt{p_z^2+2n|q_f|B}-s\mu_5)^2\right)^{3/2}}
 -\frac{1}{\left(M_{0_f}^2+p_z^2+2n|q_f|B\right)^{3/2}}\Bigg],
 \end{align}
and
 \begin{align}
 \mathbf{I}_{3,\text{log}}= \frac{N_c|q_f|B}{(2\pi)^2}\sum_{n=0}^{\infty}\sum_{s=\pm1}
 \int dp_z \frac{M_f^2}{\left(M_{0_f}^2+p_z^2+2n|q_f|B\right)^{3/2}}.
  \end{align}
 It can be shown that the term $\mathbf{I}_{3,\text{finite}}$  is finite. On the other hand the term 
 $\mathbf{I}_{3,\text{log}}$ is not convergent at large momenta. 
 Using Eq.\eqref{appenD2},
Eq.\eqref{appenD7} and Eq.\eqref{appenD9}, Eq.\eqref{appenD1} can be rearranged in the following way, 

\begin{align}
 \frac{\partial\langle\bar{\psi}_f\psi_f\rangle^{\mu_5\neq0}_{vac,B\neq0}}{\partial M_f} & = 
 \mathbf{I}_{1,\text{quad}}-\frac{M_{0_f}^2-M_f^2+2\mu_5^2}{2} \mathbf{I}_{1,\text{log}}
  +\mathbf{I}_{1,\text{finite1}}+\mathbf{I}_{1,\text{finite2}}
 +\mathbf{I}_{2,\text{finite}}+\mathbf{I}_{3,\text{finite}}\nonumber\\
& ~~~~~~~~~~~~~~~~~~~~+\mathbf{I}_{4}+\mathbf{I}_{2,\text{log}}+\mathbf{I}_{3,\text{log}}\nonumber\\
& = -\frac{M_{0_f}^2-M_f^2+2\mu_5^2}{2} \mathbf{I}_{1,\text{log}}
 +\mathbf{I}_{1,\text{finite1}}+\mathbf{I}_{1,\text{finite2}}
 +\mathbf{I}_{2,\text{finite}}+\mathbf{I}_{3,\text{finite}}\nonumber\\
& +\bigg(\mathbf{I}_{4} + \frac{N_c|q_f|B}{(2\pi)^2}\int dp_z \frac{M_f^2}{(M_{0_f}^2+p_z^2)^{3/2}}\bigg)
+(\mathbf{I}_{1,\text{quad}}+\mathbf{I}_{2,\text{log}})\nonumber\\
& +\bigg(\mathbf{I}_{3,\text{log}}-\frac{N_c|q_f|B}{(2\pi)^2}\int dp_z \frac{M_f^2}{(M_{0_f}^2+p_z^2)^{3/2}}\bigg)\nonumber\\
& = -\frac{M_{0_f}^2-M_f^2+2\mu_5^2}{2} \mathbf{I}_{1,\text{log}}
 +\mathbf{I}_{1,\text{finite1}}+\mathbf{I}_{1,\text{finite2}}
 +\mathbf{I}_{2,\text{finite}}+\mathbf{I}_{3,\text{finite}}+\mathbf{I}_{\text{finite}}+\mathbf{I}_{\text{quad}}
 +\mathbf{I}_{\text{log}},
 \label{appenD21}
 \end{align}
where  $\mathbf{I}_{\text{finite}}$ is,
\begin{align}
 \mathbf{I}_{\text{finite}} =\mathbf{I}_{4} + \frac{N_c|q_f|B}{(2\pi)^2}\int dp_z \frac{M_f^2}{(p^2+M_{0_f}^2)^{3/2}}, 
\end{align}
and
\begin{align}
\mathbf{I}_{\text{quad}}& = \mathbf{I}_{1,\text{quad}}+\mathbf{I}_{2,\text{log}}\nonumber\\
 & = -\frac{N_c|q_f|B}{2\pi^2}\Bigg[x_{0_f}(1-\ln x_{0_f})+\ln\Gamma(x_{0_f})+\frac{1}{2}\ln\left(\frac{x_{0_f}}{2\pi}\right)\Bigg]
 -\frac{2N_c}{(2\pi)^3}\int_{|\vec{p}|\leq\Lambda} d^3p\frac{1}{\sqrt{p^2+M_{0_f}^2}}.
 \label{appenD25}
\end{align}

\begin{align}
\mathbf{I}_{\text{log}}& = \mathbf{I}_{3,\text{log}}-\frac{N_c|q_f|B}{(2\pi)^2}\int dp_z \frac{M_f^2}{(p^2+M_{0_f}^2)^{3/2}},\nonumber\\
  & = -\frac{N_cM_{f}^2}{2\pi^2}\Bigg[-\ln x_{0_f}+\frac{1}{2x_{0_f}}+\frac{\Gamma^{\prime}(x_{0_f})}{\Gamma(x_{0_f})}\Bigg]
 +\frac{2N_c}{(2\pi)^3}\int_{|\vec{p}|\leq\Lambda} d^3p\frac{M_{f}^2}{\left(p^2+M_{0_f}\right)^{3/2}},
 \label{appenD26}
\end{align}
with, 
\begin{align}
\mathbf{I}_{1,\text{log}} = 
  \frac{N_c|q_f|B}{(2\pi)^2}\sum_{s=\pm 1}\sum_{n=0}^{\infty}\int dp_z \frac{1}{(p_z^2+M_{0_f}^2+2n|q_f|B)^{3/2}}
  & = -\frac{N_c}{2\pi^2}\Bigg[-\ln x_{0_f}+\frac{\Gamma^{\prime}(x_{0_f})}{\Gamma(x_{0_f})}\Bigg]\nonumber\\
 & ~~~~~~~+2N_c\int_{|\vec{p}|\leq \Lambda}\frac{d^3p}{(2\pi)^3}\frac{1}{(p^2+M_{0_f}^2)^{3/2}}.
 \label{appenD27}
\end{align}

\end{document}